\documentclass[11pt]{revtex4} 

\usepackage{amssymb}
\usepackage{amsmath}
\usepackage{verbatim}
\usepackage[pdftex]{graphicx}
\usepackage{endnotes}

\newcommand{\F}{\mathcal{F}}
\newcommand{\G}{\mathcal{G}}

\newcommand{\T}{\mathcal{T}}
\newcommand{\OO}{\mathcal{O}}
\newcommand{\HH}{\mathcal{H}}
\newcommand{\E}{\mathbb{E}}

\newcommand{\ones}{(1 \, 1 \cdots 1)^*}
\newcommand{\R}{\mathbb{R}}
\newcommand{\N}{\mathbb{N}}
\newcommand{\Z}{\mathbb{Z}}

\newcommand{\limit}[2]{\lim_{#1 \arrow #2}}
\renewcommand{\P}{P_{[0,t]}}
\newcommand{\kij}{k_{ij}}
\newcommand{\kji}{k_{ji}}
\newcommand{\deriv}[1]{\frac{\partial}{\partial #1}}
\newcommand{\utilde}{\tilde{u}_\lambda}
\newcommand{\Ltilde}{\tilde{L}_\lambda}
\newcommand{\la}{\langle}
\newcommand{\ra}{\rangle}
\newcommand{\tj}{\lfloor t_j \rfloor}
\newcommand{\arrow}{\rightarrow}

\begin{document}

\title{Fluctuation Theorems for Entropy Production and Heat Dissipation in Periodically Driven Markov Chains}

\author{Benjamin Hertz Shargel}
\affiliation{Department of Mathematics, UCLA, Los Angeles, CA, 90095-1766}
\email{shargel@math.ucla.edu}
\author{Tom Chou}
\affiliation{Departments of Mathematics and Biomathematics, UCLA, Los Angeles, CA, 90095-1766}

\begin{abstract}
Asymptotic fluctuation theorems are statements of a Gallavotti-Cohen symmetry in the rate function of either the time-averaged entropy production or heat dissipation of a process.  Such theorems have been proved for various general classes of continuous-time deterministic and stochastic processes, but always under the assumption that the forces driving the system are time independent, and often relying on the existence of a limiting ergodic distribution.  In this paper we extend the asymptotic fluctuation theorem for the first time to inhomogeneous continuous-time processes without a stationary distribution, considering specifically a finite state Markov chain driven by periodic transition rates.  We find that for both entropy production and heat dissipation, the usual Gallavotti-Cohen symmetry of the rate function is generalized to an analogous relation between the rate functions of the original process and its corresponding backward process, in which the trajectory and the driving protocol have been time-reversed.  The effect is that spontaneous positive fluctuations in the long time average of each quantity in the forward process are exponentially more likely than spontaneous {\it negative} fluctuations in the {\it backward} process, and vice-versa, revealing that the distributions of fluctuations in universes in which time moves forward and backward are related.  As an additional result, the asymptotic time-averaged entropy production is obtained as the integral of a periodic entropy production rate that generalizes the constant rate pertaining to homogeneous dynamics. 
\end{abstract}

\maketitle

\section{INTRODUCTION}

\indent Following the initial computer simulations of Evans et. al. \cite{EvansA} and the pioneering paper by Gallavotti and Cohen \cite{Gallavotti}, the study of fluctuation theorems has led to a fascinating confluence of irreversible thermodynamics, stochastic processes and large deviation theory.  Unlike its original role in statistical physics of formalizing the thermodynamic limit for equilibrium ensembles, the contribution large deviation theory makes in this context is to characterize the fluctuations in the long time average of the entropy production or heat dissipation of a stochastic process, which models a physical system whose number of degrees of freedom or incomplete description makes a deterministic treatment infeasible.  To see how this characterization comes about and these disparate fields fit together, consider an stochastic process $\xi_t$ over a state space $S$ with law $\mu(x,t)$, which for simplicity we take to be either a density or discrete distribution.  The Gibbs entropy of the process, viewed as an ensemble of paths (or a measure over that ensemble), is $-\int_S \log \mu(x,t) \mu(dx,t)$, which leads one to identify the entropy along a single stochastic trajectory as $-\log \mu(\xi_t,t)$ \cite{Seifert}.  The time derivative of this quantity equals the difference between the rate of entropy produced by the stochastic particle and the rate of entropy, or heat (divided by a nonphysical temperature), dissipated to its environment.  Maes \cite{MaesA} as well as Lebowitz and Spohn \cite{Lebowitz} further recognized that under general circumstances the total entropy production equals the logarithmic Radon-Nikodym derivative of the forward path measure $P$ governing the process with respect to its corresponding backward path measure $P^B$ which, under Crooks' more general definition \cite{Crooks}, is obtained by time-reversing all temporal inhomogeneities driving the process and composing with a path-reversal transformation.  As $P^B(\omega)$ equals the probability of observing a path $\omega$ unfold {\it in reverse} as time runs {\it backward} from time $t$ to $0$, the entropy production may be interpreted as the log likelihood of observing $\omega$ in a universe in which time runs forward as opposed to backward.  The existence of the derivative (i.e, the equivalence of $P$ and $P^B$) depends on a condition called at times dynamic reversibility \cite{MaesB} or ergodic consistency \cite{EvansB}, which ensures that the time-reversal of any trajectory realizable in the forward process is realizable in the backward process.  

The Radon-Nikodym definition for entropy production was later justified thoroughly by Maes and collaborators for a wide range of deterministic and stochastic processes \cite{MaesC,MaesB}.  It is nearly identical to that of the dissipation function $\Omega$ in deterministic mechanics \cite{EvansB,Sevick}, except that the forward and backward measures in that case put full mass on the constant energy manifold of trajectories obeying Hamilton's equations.  The logarithmic derivative is nonzero for nonstationary processes, which model physical systems evolving far from equilibrium, as well as for stationary ones violating detailed balance, modeling systems in a nonequilibrium steady state.  Its expectation under $P$ equals the relative entropy of $P$ with respect to $P^B$, which is always nonnegative, consistent with a weak reading of the second law of thermodynamics.  The time-extensive microscopic heat dissipation Lebowitz and Spohn termed an action functional, a quantity that is equal to the above logarithmic derivative up to the difference of boundary terms $-\log \frac{\mu(\xi_t,t)}{\mu(\xi_0,0)}$, precisely the net change in system entropy.  

Recent work has exploited the thermodynamic framework above by proving that either the time-averaged entropy production or heat dissipation satisfies a large deviation principle, whose corresponding rate function satisfies the same symmetry as the one proved by Gallavotti and Cohen \cite{Gallavotti} to hold for the time-averaged phase space contraction of chaotic dynamical systems.  We refer to this as an {\it asymptotic fluctuation theorem} (AFT) or, simply, a fluctuation theorem, and distinguish it from transient fluctuation theorems, which hold at finite times but are not large deviation results.  Kurchan first proved an AFT \cite{Kurchan} for the entropy production of Langevin processes under the assumption of nondegeneracy for the maximum eigenvalue of their evolution operator, and Lebowitz and Spohn then proved it \cite{Lebowitz} for the heat dissipation of time-homogeneous Markov chains and It\^{o} diffusions, whose assumed limiting stationary distribution guarantees this condition.  The cases of continuous and discrete time Markov chains were handled in a rigorous fashion by Jiang et. al. \cite{JiangBook,JiangQianZhang}, and Ge et. al. extended the discrete time case to include time-periodic inhomogeneities in the transition matrix \cite{GeJiangQian}.  Maes took a different approach from the others, studying the dynamics of finite volume Gibbs states on $\Z^{\text d}$ under general space-time potentials \cite{MaesA}.

Interestingly, it has been shown that unlike their entropy production, the heat dissipation of Langevin processes does not satisfy an AFT, at least in the conventional sense \cite{vanZonA,vanZonB}, displaying importance of the boundary terms distinguishing these quantities.  Along similar lines, R\'akos and Harris have recently shown \cite{Rakos} that infinite state spaces can result in the divergence of the boundary terms, causing a breakdown of the Gallavotti-Cohen symmetry.  It is unclear, however,  whether this breakdown, which arises from the failure of the Hamiltonian operator to satisfy Kurchan's nondegeneracy requirement, is really a function of the state space's cardinality as much as its non-compactness (the state space of a Markov chain being effectively endowed with the discrete metric).  In the case of deterministic dynamics, Bonetto et. al. have argued that apparent violations of AFTs by particle systems with singularities in the interparticle potential (owing, for example, to hard cores) can be corrected by subtracting "unphysical" singular terms from the phase space contraction rate \cite{Bonetto}.    

While so much effort recently has gone into circumscribing the range of applicability of the AFT, the purpose of our present work is to extend it for the first time to the case of inhomogeneous, continuous time dynamics.  We consider in particular a Markov chain on a finite state space, whose infinitesimal generator is a continuous and periodic function of time but only required to be irreducible at a single moment.  The finiteness of the state space ensures that the complexity of the model is isolated within the time dimension, avoiding in particular the issues raised in Ref. \cite{Rakos}.  Such a process can be used to model phenomena as diverse as the fluctuation-driven transport of molecular motors \cite{Astumian}, stochastic resonance in lasers and neuron firing \cite{Gammaitoni}, quasienergy banding in periodically-driven mesoscopic electric circuits \cite{Chen}, and seasonality in population dynamics \cite{Renshaw}, as well as periodically-driven deterministic processes amenable to coarse-graining.  Continuous time models of all of these phenomena were previously outside the scope of AFTs, all of which had been proved under the assumption of homogeneous dynamics, because they rely fundamentally on a time-dependent protocol driving the process.  Indeed, few systems in nature operate within a static environment, and so to gain true scientific relevance AFTs must ultimately accommodate time-inhomogeneities.  While our assumption of periodicity remains a restriction to potential applications, we believe that having laid out in this paper the mathematical issues involved in introducing time-dependent driving, our arguments can serve as a blueprint for future work that seeks to loosen this restriction.

Whereas the authors mentioned above have proved AFTs for either the action functional or entropy production of a process alone, we prove for both, finding that the absence of boundary terms in the action functional simplifies the derivation of its free energy but complicates that of its fluctuation symmetry.  The fluctuation symmetry for both quantities takes the form $I(z) - I^B(-z) = z$, where $I$ is the rate function under the forward process and $I^B$ under the backward process, which reduces to the usual Gallavotti-Cohen symmetry when the driving protocol is symmetric inside every driving period and, hence, has no temporal orientation.  Its interpretation is that spontaneous positive fluctuations in the long time average of each quantity in one process is exponentially more likely than spontaneous {\it negative} fluctuations in the other, a relation that is symmetric with respect to the two processes.

We also derive the almost sure asymptotic time-averaged entropy production as the integral $\int_0^T e_p(s)ds$, where $e_p(s)$ is a $T$-periodic instantaneous entropy production rate, with $T \in \R^+$ the period of the driving.  This expression generalizes existing ones known for homogeneous \cite{Gaspard,JiangQianZhang} and periodically inhomogeneous, but discrete time \cite{GeJiangQian}, chains.  Our proofs are guided strategically by those of Lebowitz and Spohn, but, as discussed above, the inhomogeneity of the process necessitates more involved, and rigorous, arguments.  The virtue of this is that our proofs, unlike those of Leibowitz and Spohn, do not rely explicitly on the existence of a stationary distribution for the process (indeed, only a periodic quasi-stationary distribution exists), raising the open question of how insensitive the existence of an AFT is to the details of the asymptotic regime of the process.  Put another way, how far can our assumption of periodicity on the driving rates be relaxed but still guarentee an AFT?  We address this question in the final section of the paper, proving that uniform continuity and boundedness of the rates alone, which are implied by the assumptions described in section II, are not sufficient.  This shows that any minimal set of conditions on the rates are closer to those assumed in this paper than one might initially suspect.

A number of theoretical and experimental results already exist for periodically-driven stochastic processes.  Integral and transient fluctuation theorems have been proved for periodically-driven two-state Markov chains \cite{Schuler,Tietz}, quantum systems \cite{KurchanB}, as well as classical harmonic oscillators modeled by a Langevin process \cite{Joubaud,Singh}.  It must be emphasized that these studies differ from ours because they do not prove large deviations results.  As the technicalities we deal with in this paper reveal, the transition from transient to asymptotic fluctuation theorem, even for processes as simple as finite state Markov chains, is not an automatic or obvious one, but depends on the details of the driving protocol.

Similarly, in spite of the close relation between discrete and continuous time Markov chains, our results cannot be obtained from those of Ge. et. al.  The set of transition probabilities in their case is finite, and the uniformity of the waiting times between jumps enables the inhomogeneous chain on $S$ to be re-represented as a homogeneous chain on the larger (but still finite) space $S^{T'}$, where $T' \in \N$ is the period of the transition probabilities.  Our set of transition rates is uncountable, on the contrary, and is sampled randomly by the process, ensuring no convenient reduction to a homogeneous problem.  It is possible, of course, that with the proper scaling of the transition probabilities (under which the periodicity $T'$ would diverge), the continuous time path measure may be obtained as the weak limit of the discrete time path measures, analogous to Donsker's theorem for Brownian motion.  However, as has been mentioned, the rate functions for the heat dissipation and entropy production of the continuous process do not simply inherit the fluctuation symmetry of their discrete approximants.  Instead, the internal symmetry of each rate function is replaced by a symmetry between it and its counterpart under the corresponding backward process.

The rest of the paper is organized as follows.  We begin by making the appropriate definitions, and then derive a backward equation for the moment generating function of the action functional, whose fundamental solution we obtain using both Floquet theory and the time-ordered exponential operator.  This enables us to identify the free energy of the action functional as the nondegenerate principal Floquet eigenvalue of the fundamental solution, thereby proving the existence and strict convexity of its Legendre-Fenchel transform, the rate function.  We then show that the free energies of the action functional under the forward and backward path measures satisfy a symmetry relation analogous to the one found by Lebowitz and Spohn in the case of time-homogeneous dynamics, implying the fluctuation theorem symmetry between the forward and backward rate functions.  All of these results are subsequently proved for the entropy production in place of the action functional, along with the consequent generalization of the second law of thermodynamics and a derivation of the asymptotic time-averaged entropy production and associated production rate.  We conclude with a look beyond periodic driving, considering the case of uniformly continuous and bounded rates. 

\section{DEFINITIONS AND SETUP}

Let $(X_t)_{t \geq 0}$ be a continuous time Markov chain on the probability space $(\Omega, \F, P)$, where $\Omega=D(0,\infty)$ is the space of c\`adl\`ag paths (i.e., right-continuous with left limits) over a finite state space, which, without loss of generality, we take to be $\{1, 2, \ldots, N\}$, and $P$ is the Markov path measure with initial distribution $\pi$.  The process generates the right-continuous filtration $\F_t = \bigcap_{s>t} \, \sigma(X_r, 0 \leq r \leq s)$, representing information known about the process up to and infinitesimally beyond time $t$, with $\F_t \uparrow \F$, and evolves according to transition rates $\kij(t)$, such that the probability of jumping from state $i$ to $j$ within a time window $[t,t+\tau]$ equals $\kij(t) + o(\tau)$.  These rates are assumed to be such that the infinitesimal generator $A(t) = (\kij(t))_{i,j=1}^N$ is continuous $\forall t \geq 0$, $T$-periodic, and irreducible for some $t^* \in [0,T]$, but whose adjacency graph, representing which states currently communicate, may otherwise change over time and become reducible.  We further require that the rates satisfy the so-called dynamic reversibility \cite{MaesB} or ergodic consistency condition \cite{EvansB}, whereby $\kij(t)>0 \iff \kji(t)>0$, which again ensures that the time-reversal of any realizable trajectory in the forward process is realizable in the backward one.  The law of the process, $\mu(i,t) = P(\omega \in \Omega: X_t(\omega) = i)$, satisfies the forward Kolmogorov equation \cite{Lebowitz}

\begin{equation}
\frac{\partial \mu}{\partial t} = A^*(t) \mu
\label{FORWARD}
\end{equation}

\noindent with initial condition $\mu(\cdot,0)=\pi$, where $\cdot^*$ denotes the adjoint.  We take $(\tau_i)_{i \geq 1}$ to be the random jumping times of the process and $\sigma_i = X_{\tau_i}$, with $\sigma_0 = X_0$.  

Let us now define the objects we will primarily be concerned with, the entropy production

\begin{equation*}
S(t_0,t) = \log \frac{dP_{[t_0,t]}}{dP_{[0,t-t_0]}^B},
\end{equation*}

\noindent equal to the logarithmic Radon-Nikodym derivative between the forward and backward path measures (to be discussed shortly), and the action functional

\begin{equation}
W(t_0, t) = S(t_0,t) - \log \frac{\mu(X_{t_0},t_0)}{\mu(X_t,t)}
= \sum_{t_0 < \tau_i < t} \log 
	\frac{k_{\sigma_{i-1}\sigma_i}(\tau_i)}{k_{\sigma_i\sigma_{i-1}}(\tau_i)}
\label{AF}
\end{equation}

\noindent representing heat dissipation, where $\log (\mu(X_{t_0},t_0)/ \mu(X_t,t))$ equals the net difference in system entropy between times $t_0$ and $t$. ($\mu(X_s,s)$ here should be interpreted as $\mu(i,s)|_{i=X_s(\omega)}$.)  Exponential factors representing holding times between jumps have been canceled on the RHS of (\ref{AF}) (see Refs. \cite{GeJiang,Harris}); a full representation of the forward path density can be found in (\ref{DENSITY}) in Section V.  For a comprehensive justification of the definitions used above for entropy production and heat dissipation, see Refs. \cite{Harris,JiangBook,MaesB,MaesC,Seifert}.

$P_{[t_0,t]}$ above is our Markov path measure restricted to $[t_0,t]$ and $P_{[0,t-t_0]}^B$ is the corresponding backward path measure, defined as follows.  Let $P_{[0,t-t_0]}^-$ be a measure obtained from $P_{[t_0,t]}$ by setting its initial distribution $\mu^-(\cdot,0) = \mu(\cdot,t)$ and its rates $\kij^-(s) = \kij(t-s)$ for $0 \leq s \leq t-t_0$, and let $r(\omega)_s = \lim_{s' \uparrow t - s} \, \omega_{s'}$ be the path-reversal transformation, with $\omega \in \Omega$ a sample path.  We then define $P_{[0,t-t_0]}^B \equiv P_{[0,t-t_0]}^- \circ r$.  Note that $r$ is involutive on $\Omega$, preserving the c\`adl\`ag property of paths, and that the law of the backward process, $\mu^B(\cdot,s)$, defined for $0 \leq s \leq t-t_0$, is implicitly a function of $t$ due to $r$.  It satisfies the final condition $\mu^B(\cdot,t) = \mu^-(\cdot,0)=$ $\mu(\cdot,t)$.  Note crucially that the backward process is not a Markov process with respect to the filtration $(\F_s)_{0 \leq s \leq t}$.  Indeed, $\mu^B(\cdot,0) = \mu^-(\cdot,t)$, and so the smallest $\sigma$-algebra that the event $\{X_0 = i\}$ is measurable with respect to under this process is $\F_t$.  It is, however, Markovian with respect to the backward time filtration $\G_t = \bigcap_{s<t} \, \sigma(X_r, s \leq r)$.

Reversing both the path and the rates in the backward path measure ensures that the transitions of a reversed path in the $P^-$-governed process occur under the same local conditions (i.e., instantaneous rates) as those of the original path do in the forward process.  Intuitively, if $\P(\omega)$ is the probability of observing the trajectory $\omega$ unfold as time runs forward from $0$ to $t$, then $\P^B(\omega)$ is the probability of observing that trajectory unfold {\it in reverse} as time runs {\it backward} from $t$ to $0$, with the filtration $\G_t$ representing the past.  From an Archimedian perspective "outside of time" \cite{Price}, neither direction of time should be preferred {\it a priori}.  $S(0,t)$ is therefore truly a measure of irreversibility: Thinking of $\omega$ as a spacetime curve instead of an oriented trajectory, $S(0,t)(\omega)$ gives the log likelihood of $\omega$ being realized in a universe where time runs forward as opposed one in which time runs backward, with the sign of its expectation - dependent on which measure we integrate with respect to - indicating the overall direction of time's arrow.  

It has been argued recently \cite{Feng}, we should note, that the expectation of $S(0,t)$, the mutual entropy $\HH(\P,\P^B)$, is too sensitive to rare irreversible events to be a useful measure of irreversibility, and that the quantity $A = \frac{1}{2} \HH(\P,\frac{1}{2}(\P + \P^B)) + \frac{1}{2} \HH(\P^B,\frac{1}{2}(\P + \P^B))$ should be used instead.  The "asymmetry" $A$ equals the amount of information gained about the direction of time's arrow from watching one realization of the process.

The first theorem that we will prove is the existence and differentiability of the free energy function of the action functional $W(0,t)$,

\begin{equation}
c_W(\lambda) = \limit{t}{\infty} \frac{1}{t} \, 
	\log \E_{\pi,0} \bigl(e^{\lambda W(0,t)} \bigl),
\label{FREE_ENERGY}
\end{equation}

\noindent which guarantees a large deviation property for $W(0,t)/t$ with rate function $I_W(z) = \sup_{\lambda \in \R} \{\lambda z - c_W(\lambda) \}$.  Here $\E_{\nu,t_0}(\cdot)$ denotes expectation conditioned on the chain having distribution $\nu$ at time $t_0$.  Our second main result is the symmetry relation $c_W(\lambda) = c_W^B(-(1+\lambda))$ between it and the free energy of the action functional under the backward process,

\begin{equation}
c_W^B(\lambda) = \limit{t}{\infty} \frac{1}{t} 
	\, \log \E_{\mu(\cdot,t),t}^B \bigl(e^{\lambda W^B(0,t)} \bigl),
\label{BACKWARD_FREE_ENERGY}
\end{equation} 

\noindent which reduces to the usual Gallavotti-Cohen symmetry when the generator is a symmetric function of time within each driving period, in which case the backward path measure reduces to the path-reversed measure $\P \circ r$ considered in Refs. \cite{JiangBook,JiangQianZhang,Lebowitz,MaesC,MaesB}.  (Note that even in this scenario these authors' results do not apply, because the dynamics remain time-dependent and non-stationary.)  See Ref. \cite{Schuler} for a discussion in the context of transient fluctuation theorems. 

In the definition (\ref{BACKWARD_FREE_ENERGY}), the expectation is taken with respect to $\P^B$, and $W^B(0,t)$ represents the heat dissipated by a process that traces the curve $\omega$ backward through time.  Since each transition $\sigma_i \arrow \sigma_{i-1}$ at time $\tau_i$ increases $W^B(0,t)$ by $\log k_{\sigma_i\sigma_{i-1}}(\tau_i)/ k_{\sigma_{i-1}\sigma_i}(\tau_i)$, we simply have $W^B(0,t) = -W(0,t)$.  Similarly, the entropy production under the backward process $S^B(0,t)= \log d\P^B/d\P = -S(0,t)$.  The key point is that while $W$ and $W^B$ (resp. $S$ and $S^B$) are distinct functions, when evaluated under $\P$ and $\P^B$, respectively, they represent the {\it same physical quantity}.  For additional discussion of this and the relation between forward and backward measures in general, see section 3.1 of Ref. \cite{Harris}.

\section{LARGE DEVIATIONS OF HEAT DISSIPATION}

To prove the existence of the limit (\ref{FREE_ENERGY}), we first derive a backward equation for the time-dependent moment generating function $u_\lambda(t_0,t)$, whose elements $u_\lambda(i,t_0,t)$ equal $\E_{i,t_0} [e^{\lambda W(t_0,t)}]$, where the expectation is conditioned on the process being at state $i$ at time $t_0$.  For notational economy we define $\Lambda_\lambda(t_0,t) = e^{\lambda W(t_0,t)}$ and $w(i,j,t) = \log (\kij(t) / \kji(t))$.  As a general rule we use $\lambda$ as a subscript in order to emphasize its distinction from time parameters, with the exception of the free energy.  Finally, we use the convention that all matrix inequalities, denoted by $\preceq$, and limits are defined component-wise.

\bigskip

{\bf Proposition 3.1:} {\it $u_\lambda(t_0,t)$ satisfies the Kolmogorov backward equation}

\begin{equation}
\deriv{t_0} u_\lambda(t_0,t) = -L_\lambda(t_0) \, u_\lambda(t_0,t)
\label{BACKWARD}
\end{equation}

\noindent {\it where}

\begin{equation}
L_\lambda(t_0)_{i,j} = 
\begin{cases}
\kij(t_0)^{1+\lambda} \kji(t_0)^{-\lambda}, & i \neq j \\
-K_i(t_0), & i=j,
\end{cases}
\label{L_DEF}
\end{equation} 

\noindent {\it and $K_i(t_0) = \sum_{j \neq i}^N \kij(t_0)$ is the escape rate from state $i$ at time $t_0$.}

\bigskip

{\bf Proof:}  Because the jump times of a Markov chain are isolated almost surely and its paths are c\`adl\`ag, we interpret the partial derivative in (\ref{BACKWARD}) to be the left derivative: 
$\deriv{t_0} u_\lambda(t_0,t) = \limit{h}{0+} 
\frac{u_\lambda(t_0,t) - u_\lambda(t_0 - h,t)}{h}$.
Evaluating the second term in the numerator is the crux of the problem.  We begin with

\begin{align*}
\limit{h}{0+} u_\lambda(i,t_0 - h,t) 
		   & = \limit{h}{0+} \E_{i,t_0 - h} [\Lambda_\lambda(t_0 - h, t)] \\
		   & = \limit{h}{0+} \E_{i,t_0 - h} 
				[ \E_{i,t_0-h}(\Lambda_\lambda(t_0-h, t_0) \Lambda_\lambda(t_0+, t) | 
				\F_{t_0})] \\
		   & = \limit{h}{0+} \E_{i,t_0 - h} 
				[\Lambda_\lambda(t_0-h, t_0) \E_{i,t_0-h}(\Lambda_\lambda(t_0+, t) | 
				\F_{t_0})] \\
		   & = \limit{h}{0+} \E_{i,t_0 - h} 
				[\Lambda_\lambda(t_0-h, t_0) \E_{X_{t_0},t_0}(\Lambda_\lambda(t_0+, t)].
\end{align*}		   

\noindent Here the limit $t_0+$ is taken before the limit $h \downarrow 0$, and in the last line we have used the Markov property inherited by $\Lambda_\lambda(t_0,t)$ from $X_t$.  Since $t_0$ is not an accumulation point for jumps almost surely and we are operating in the small $h$ limit, at most one jump can occur between $t_0-h$ and $t_0$.  The last line therefore becomes

\begin{align*}
& \limit{h}{0+} \E_{i,t_0 - h} \bigl[ e^{\lambda w(i,X_{t_0},t_0)} \,
	\E_{X_{t_0},t_0} \bigl( \Lambda_\lambda(t_0+, t) \bigl) \bigl] \\
& = \limit{h}{0+} \sum_{j=1}^N \E_{i,t_0 - h} \bigl[ e^{\lambda w(i,X_{t_0},t_0)} \,
	\E_{X_{t_0},t_0} \bigl( \Lambda_\lambda(t_0+, t) \bigl) 1_{X_{t_0}=j} \bigl] \\
& = \limit{h}{0+} \sum_{j=1}^N e^{\lambda w(i,j,t_0)} \,
	\E_{i,t_0 - h} [\E_{j,t_0} \bigl( \Lambda_\lambda(t_0+, t) \bigl) 1_{X_{t_0}=j}],
\end{align*}

\noindent with $1_A$ denoting the indicator of the event $A$.  Regardless of whether a jump occurs at $t_0$ (that is, $X_{t_0} = j \neq i$), no jumps may occur during a sufficiently small period after $t_0$.  The inner expectation therefore equals $u_\lambda(j,t_0,t)$ and our limit is

\begin{align}
& \limit{h}{0+} \sum_{j=1}^N e^{\lambda w(i,j,t_0)} u_\lambda(j,t_0,t)   
	\E_{i,t_0 - h}[1_{X_{t_0}=j}] \notag \\
\label{BACKWARD_LINE}
& = \limit{h}{0+} \sum_{j=1}^N e^{\lambda w(i,j,t_0)} u_\lambda(j,t_0,t)
	p(i,j,t_0-h,t_0),
\end{align}

\noindent where $p(i,j,t_0-h,t)$ is the transition probability of being at $j$ at time $t_0$ given having been at $i$ at $t_0-h$.  For $i \neq j$ this equals $\kij(t_0-h) h + o(h)$ and, for $i=j$, it equals $1 - K_i(t_0-h)h + o(h)$, with $K_i$ defined in the statement of the Proposition.  We therefore have, by continuity of the $\kij$,

\begin{equation*}
\limit{h}{0+} u_\lambda(i,t_0 - h,t)
= \limit{h}{0+} \sum_{j \neq i}^N (\kij(t_0) h + o(h)) e^{\lambda w(i,j,t_0)} 
u_\lambda(j,t_0,t) + (1 - K_i(t_0)h + o(h)) u_\lambda(i,t_0,t)
\end{equation*}

\noindent and the component-wise derivative becomes

\begin{align*}
\deriv{t_0} u_\lambda(i,t_0,t) 
& = \limit{h}{0+} \frac{1}{h} [u_\lambda(i,t_0,t) - u_\lambda(i,t_0 - h,t)] \\
& = - \biggl[ \sum_{j \neq i}^N \kij(t_0)^{1+\lambda} \kji(t_0)^{-\lambda} \, 
	   u_\lambda(j,t_0,t) - K_i(t_0) u_\lambda(i,t_0,t) \biggl] \\
\end{align*}

\noindent Collecting components into a vector equation, we obtain (\ref{BACKWARD}).  $\Box$

\bigskip

\noindent Before stating the first main theorem, we introduce a concept from quantum field theory that will be useful.  For a family of operators $\{\OO(s)\}_{s \in \R}$, define the time-ordering operator $\T$ \cite{Sterman} by $\T \prod_{j=0}^n \OO(s_j) = \OO(s_n)\cdots \OO(s_1)\OO(s_0)$, where $s_0 < \cdots < s_n$ is any finite sequence in $\R$.  
Now, given $t_0 < t_1$, we define the time-ordered exponential

\begin{equation*}
\T \biggl( \exp \int_{t_0}^{t_1} \OO(s) ds \biggl) \,
= \limit{n}{\infty} \exp \biggl(\frac{t_1 - t_0}{n} \OO(s_{n-1}^n) \biggl)
			\cdots \exp \biggl(\frac{t_1 - t_0}{n} \OO(s_1^n) \biggl)
			\exp \biggl(\frac{t_1 - t_0}{n} \OO(s_0^n) \biggl)
\end{equation*}

\noindent when the limit exists, where $\{s_j^n = t_0+j(t_1-t_0)/n\}_{j=0}^n$ is a sequence of uniform partitions of $[t_0,t_1]$.  

Recall that a matrix is called {\it quasipositive} if its entries are nonnegative, strictly on the diagonal.  Our results rely on the fact that the time-ordered exponential operator, like the usual exponential operator, transforms quasipositivity into strict positivity.  The proof requires some clever bounding arguments.

\bigskip

{\bf Lemma 3.2:} {\it Let $0 \leq t_0 < t_1$ and $L: [0,\infty) \arrow M_{N \times N}(\R)$ be a continuous function whose values are quasipositive $\forall t \geq 0$ and irreducible for some $t^* \in (t_0, t_1)$.  Then the matrix 
$\T ( \exp \int_{t_0}^{t_1} L(t) dt )$ has strictly positive entries.}

\bigskip

{\bf Proof:}  Without loss of generality, we take $t_0 = 0$, the length of the interval being the only relevant factor.  By uniform continuity of $L(t)$ on $[0,t_1]$, choose $\sup_{t \in [0,t_1], 1 \leq i \leq N} L(t)_{ii}^- < M < \infty$, where the superscript denotes the negative part of the number, so that $E(t) = L(t) + MI$ has strictly positive diagonal entries and nonnegative off-diagonal ones, $\forall t \in [0,t_1]$.  Employing the partitions $\{s_j^n\}$ defined above, we have

\begin{align}
& \exp \biggl(\frac{t_1}{n}L(s_{n-1}^n) \biggl) \cdots 
	\exp \biggl(\frac{t_1}{n}L(s_0^n) \biggl) \notag \\
& \quad = \exp \biggl(\frac{t_1}{n}L(s_{n-1}^n) 
	+ \frac{M t_1}{n}I - \frac{M t_1}{n}I \biggl) \cdots 
	\exp \biggl(\frac{t_1}{n}L(s_0^n) + \frac{M t_1}{n}I 
	- \frac{M t_1}{n}I \biggl) \notag \\
& \quad = e^{-M t_1} \exp \biggl(\frac{t_1}{n}E(s_{n-1}^n) \biggl) 	
	\cdots \exp \biggl( \frac{t_1}{n} E(s_0^n) \biggl) \notag \\
\label{EXPONENTIALS}
& \quad \succeq e^{-M t_1} \biggl( I + \frac{t_1}{n} E(s_{n-1}^n)
	\biggl) \cdots \biggl( I + \frac{t_1}{n} E(s_0^n) \biggl).
\end{align}

\noindent By irreducibility of $L(t^*)$, let $0 < \alpha < \min_{1 \leq i,j \leq N} (E(t^*)^N)_{i,j}$, and define $E_\alpha$ to be the $N \times N$ matrix whose entries are all $\alpha$.  By continuity of $E(t)$, choose $\delta$ small enough so that $\forall t^* < u_1, \dots, u_N < t^* + \delta$, $E(u_1) \cdots E(u_N) \succeq E_\alpha$.  Finally, let $k = \lceil nt^* / t_1 \rceil$ and $\ell = \lfloor n(t^*+\delta) / t_1 \rfloor$ denote the index of the first partition point after $t^*$ and the last one before $t^*+\delta$, respectively, so that $\ell - k \geq n \delta /t_1 - 2$.  We can then rewrite (\ref{EXPONENTIALS}) as 

\begin{align*}
& e^{-M s_1} \biggl( I + \frac{t_1}{n} E(s_{n-1}^n) \biggl) \cdots
	\biggl( I + \frac{t_1}{n} E(s_{\ell+1}^n) \biggl)
	\biggl( I + \frac{t_1}{n} E(s_\ell^n) \biggl) \cdots
	\biggl( I + \frac{t_1}{n} E(s_k^n) \biggl) \\
& \quad \times 
	\biggl( I + \frac{t_1}{n} E(s_{k-1}^n) \biggl) \cdots
	\biggl( I + \frac{t_1}{n} E(s_0^n) \biggl) \\
& \quad \succeq e^{-M t_1} \biggl( I + \frac{t_1}{n} E(s_\ell^n) \biggl)
	\cdots \biggl( I + \frac{t_1}{n} E(s_k^n) \biggl) \\
& \quad \succeq e^{-M t_1} {{\ell - k} \choose N} 
	\biggl(\frac{t_1}{n} \biggl)^N E_\alpha \\
& \quad \succeq e^{-Mt_1} 
	\frac{(n\delta/t_1 - 2) \cdots (n\delta/t_1 - (N+1))}{(n/t_1)^N}
	\frac{1}{N!} E_\alpha \\
& \quad \xrightarrow{n \arrow \infty} e^{-Mt_1} \frac{\delta^N}{N!} E_\alpha.
\end{align*}

\noindent Therefore, choosing $0 < \beta < e^{-M t_1} \delta^N \alpha /N!$, we have that the entries of $\T ( \exp \int_0^{t_1} L(t) dt )$ are bounded below by $\beta$.   $\Box$

\bigskip

\noindent Note that by the irreducibility and ergodic consistency conditions on $A(t)$, $L_\lambda(t)$ is irreducible at $t^*$, and, hence, the lemma applies to it.  We recall that our assumptions on the generator $A(t)$ are sufficiently weak that the rates $\kij(t)$ may drop down to zero, thereby changing the structure of its adjacency graph and possibly rendering it reducible.  Irreducibility of $A(t)$ over an arbitrarily small interval is sufficient, nevertheless, for all states of the chain to communicate after $t^*$.

\bigskip

{\bf Theorem 3.3:} {\it Given the periodicity, continuity, irreducibility and ergodic consistency conditions on the generator $A(t)$, the free energy $c_W(\lambda)$ exists, is continuously differentiable $\forall \lambda \in \R$, and is independent of $\pi$.}

\bigskip

{\bf Proof:}  Rewriting the free energy as 
$\limit{t}{\infty} \frac{1}{t} \, \log \la \pi,u_\lambda(0,t) \ra$, where $\la \cdot \ra$ denotes the inner-product on $\R^{\text {N}}$, our strategy is to represent $u_\lambda(0,t)$ in terms of the Floquet fundamental solution to (\ref{BACKWARD}), whose associated flow operator we will show has strictly positive entries and whose periodic component can be bounded so as not to affect the asymptotics.  The Perron-Frobenius theorem will then allow us to identify the free energy as the principal Floquet eigenvalue of the fundamental solution, which depends smoothly on $\lambda$.  

We begin by noticing that in our backward equation, $u_\lambda(t_0,t)$ evolves according to the matrix $-L_\lambda(t_0)$, whose off-diagonal terms are non-positive and whose corresponding flow will therefore not have exclusively positive entries.  Instead of integrating forward, we therefore elect to change variables and integrate {\it back} to $u_\lambda(0,t)$ from the known value $u_\lambda(t,t) \equiv \ones$.  To wit, let $\tau = t - t_0$, and define $\utilde(\tau,t)=u_\lambda(t-\tau,t)=u_\lambda(t_0,t)$ and similarly $\Ltilde(\tau) = L_\lambda(t_0)$.
Then

\begin{equation}
\deriv{\tau} \utilde(\tau,t) = - \deriv{t_0} u_\lambda(t_0,t) 
= L_\lambda(t_0) u_\lambda(t_0,t) = \Ltilde(\tau) \utilde(\tau,t).
\label{BACKWARD2}
\end{equation}

Recall that the fundamental solution to (\ref{BACKWARD2}) with initial condition $\utilde(0,t)=\ones$ is a square matrix whose $N$ columns are $N$ linearly independent solutions to the ODE.  The Floquet theorem \cite{Verhulst} guarantees that by periodicity of $\Ltilde(\tau)$, this fundamental solution has the form $\Phi_\lambda(\tau,t) = e^{\tau H_\lambda(t)}P_\lambda(\tau,t)$, where $P_\lambda$ is $T$-periodic and continuous in $\tau$ and $H_\lambda(t)$ is complex.  We choose in particular the principle fundamental solution, which satisfies $\Phi_\lambda(0,t)=I$, so that $\utilde(\tau,t) = \Phi_\lambda(\tau,t)\utilde(0,t)$.  This theorem is derived by first noting that by periodicity of $\Ltilde$, if $\Phi_\lambda(\tau,t)$ is a fundamental solution of (\ref{BACKWARD2}), then so is $\Phi_\lambda(\tau+T,t)$.  This implies that the two matrices are linearly dependent, and hence there exists a nonsingular matrix $C(t)$ such that $\Phi_\lambda(\tau+T,t) = C(t)\Phi_\lambda(\tau,t)$.  $H_\lambda(t)$ is defined as $\frac{1}{T}\log C(t)$, since nonsingular matrices always possess a logarithm.  We may therefore identify $e^{T H_\lambda(t)}$ as the flow operator that evolves solutions $T$ units of time into the future, which is {\it independent} of the current time $\tau$.  From (\ref{BACKWARD2}), it can then be represented as

\begin{equation}
e^{T H_\lambda(t)} = \T \biggl( \exp \int_\tau^{\tau+T} \Ltilde(s) ds \biggl),
\quad \forall \, 0 \leq \tau \leq t-T,
\end{equation}

\noindent which, by the lemma, has strictly positive entries.  A second dividend of this representation is that because the RHS does not depend on $t$, neither does $H_\lambda$.  

By the Perron-Frobenius theorem, $e^{T H_\lambda}$ possesses a nondegenerate positive, maximum eigenvalue $e^{T \vartheta(\lambda)}$ and a positive eigenvector $v_\lambda$ spanning its corresponding one-dimensional eigenspace.  This makes $\vartheta(\lambda)$ the nondegenerate principal Floquet eigenvalue of the fundamental solution $\Phi_\lambda$, which, as a simple root of the characteristic equation of $H_\lambda$, is continuously differentiable with respect to $\lambda$ by the implicit function theorem \cite{Lax}.  Evaluating the free energy function,

\begin{align}
\begin{split}
c_W(\lambda) & = \limit{t}{\infty} \frac{1}{t} \log \E_{\pi,0} 
				\bigl( e^{\lambda W(0,t)} \bigl) \\
	& = \limit{t}{\infty} \frac{1}{t} \log \, \la \pi, u_\lambda(0,t) \ra \\
	& = \limit{t}{\infty} \frac{1}{t} \log \, \la \pi, \utilde(t,t) \ra \\
	& = \limit{t}{\infty} \frac{1}{t} \log \, \la \pi, 
		e^{t H_\lambda} P_\lambda(t,t) \ones \ra \\
	& = \limit{t}{\infty} \frac{1}{t} \log \, \bigl( \la \pi, v_\lambda\ra
		e^{t \vartheta(\lambda)} 
		\la v_\lambda, P_\lambda(t,t) \ones \ra \bigl) \\
	& = \vartheta(\lambda) + 
		\limit{t}{\infty} \frac{1}{t} \log \, 
		\la v_\lambda, P_\lambda(t,t) \ones \ra
\end{split}
\label{CHAIN}
\end{align}		   

\noindent where we have used the uniqueness of the maximal eigenvalue and then the positivity of $v_\lambda$ to eliminate the dependence on $\pi$.  

Our final task is to show the remaining limit vanishes by bounding the quadratic form $Q(t) = \la v_\lambda, P_\lambda(t,t)\ones \ra$ between positive numbers, uniformly in $t$.  To achieve this, we again exploit the time-ordered exponential representation of the flow.  Taking $t = mT + r$, where $0 \leq r < T$, by the factoring property of such operators \cite{Sterman} 

\begin{align*}
e^{mT H_\lambda} e^{r H_\lambda} P_\lambda(t,t)
& = \T \biggl( \exp \int_0^t \Ltilde(s)ds \biggl) \\
& = \T \biggl( \exp \int_r^t \Ltilde(s)ds \biggl) 
	\T \biggl( \exp \int_0^r \Ltilde(s)ds \biggl) \\
& = e^{mT H_\lambda} \T \biggl( \exp \int_0^r \Ltilde(s)ds \biggl). \\
\end{align*}

\noindent Multiplying by $e^{-r H_\lambda} e^{-mT H_\lambda}$ on the left, we have 

\begin{equation}
P_\lambda(t,t) = e^{-r H_\lambda} \T \biggl( \exp \int_0^r \Ltilde(s)ds \biggl).
\label{P_LAMBDA}
\end{equation}

\noindent The presence of $r$ but not $t$ on the RHS reveals that $\bar{P}_\lambda(t) \equiv P_\lambda(t,t)$ is $T$-periodic, and the form of (\ref{P_LAMBDA}) ensures that it is uniformly continuous on its periodic domain.  We may conclude from this that the matrix norm of $P_\lambda(t,t)$ is bounded in $t$, and so the quadratic form is bounded above.  

Because we may have $r<t^*$, the time at which the generator is irreducible, the lemma does not guarantee for us that $\T (\exp \int_0^r \Ltilde(s)ds)$ has strictly positive entries.  However, it is straightforward to see by adding and subtracting $M > \sup_{s \in [0,r], 1 \leq i \leq N} \Ltilde(s)_{ii}^-$ times the identity from $\Ltilde(s)$ that this matrix exponential does have nonnegative entries, strictly on the diagonal.  From this we may conclude that

\begin{equation*}
w_\lambda \equiv \T \biggl( \exp \int_0^r \Ltilde(s)ds \biggl) \ones \succeq 0
\end{equation*}

\noindent with $w_\lambda \neq 0$.  Taking $R_{v_\lambda^\perp}$ to be the projection operator onto the orthogonal subspace of $v_\lambda$, by (\ref{P_LAMBDA}),

\begin{align*}
Q(t) & = \la v_\lambda, P_\lambda(t,t) \ones \ra \\
	 & = \la v_\lambda, e^{-r H_\lambda} \T \biggl( \exp \int_0^r \Ltilde(s)ds \biggl) 
			\ones \ra \\
	 & = \la v_\lambda, e^{-r H_\lambda} w_\lambda \ra \\
	 & = \la v_\lambda, e^{-r H_\lambda} 
			(\la v_\lambda,w_\lambda \ra v_\lambda + R_{v_\lambda^\perp} w_\lambda) \ra \\
	 & = e^{-r \vartheta(\lambda)} \la v_\lambda, w_\lambda \ra^2 > 0.
\end{align*}

\noindent This is our uniform lower bound.   $\Box$

\bigskip

{\bf Remark 3.4:} One novelty of the proof of Theorem $3.3$ vis-\'a-vis the proofs of large deviation principles by Lebowitz and Spohn \cite{Lebowitz}, Jiang et. al. \cite{JiangQianZhang} and others is that it does not rely on the existence of an ergodic distribution for the dynamics.  Indeed, no stationary distribution exists.  By (\ref{FORWARD}), $\mu(\cdot,t)$ evolves according to the periodic matrix $A^*(t)$, and so by the Floquet theorem can be represented as $\mu(\cdot,t) = e^{tB}P(t)\pi$, where $P(t)$ is $T$-periodic.  Because $A^*(t)$ is quasipositive always and irreducible at $t^*$, by Lemma $3.2$ and an argument analogous to that in the proof of Theorem $3.3$, $e^{tB}$ has strictly positive entries and thus a Perron-Frobenius eigenvalue.  This eigenvalue must be $1$, otherwise probability conservation would be violated.  Taking $v$ to be the corresponding positive eigenvector and $R_{v^\bot}$ the projection onto the subspace of $\R^N$ orthogonal to it,

\begin{equation}
\label{ASYMPTOTIC}
\mu(\cdot,t) = e^{tB}P(t)\pi = \la P(t)\pi,v \ra v  + e^{tB} R_{v^\bot} P(t)\pi 
\sim \la P(t)\pi,v \ra v.
\end{equation}

\noindent We see that the asymptotic limit of {\it any} initial distribution $\pi$ is $T$-periodic through $P(t)$, and therefore that no stationary distribution exists.  As proved in Theorem 5.2, this asymptotic limit takes the place of the usual stationary distribution in the expression for the instantaneous entropy production rate.

\bigskip

{\bf Corollary 3.5:} {\it The time-averaged action functional $W(0,t)/t$ satisfies a large deviation principle with continuous and strictly convex good rate function} 

\begin{equation}
\label{W_TRANSFORM}
I_W(z) = \sup_{\lambda \in \R} \{\lambda z - c_W(\lambda) \}.
\end{equation}

{\bf Proof:} Existence and differentiability of $c_W(\lambda)$ everywhere imply, by the G\"{a}rtner-Ellis theorem \cite{EllisA}, that $W(0,t)/t$ satisfies a large deviation principle with rate function $I_W(z) = \sup_{\lambda \in \R} \{\lambda z - c_W(\lambda) \}$.  Strict convexity of $I_W(z)$ can be deduced by contradiction.  If the convex envelope of $I_W(z)$ (the supremum of all convex functions minorizing it) contained an interval of strict linearity with slope $\lambda_0$, then the free energy's derivative would jump at $\lambda_0$ by Legendre duality, violating its continuity (see Ref. \cite{Touchette} for an intuitive discussion).

Continuity requires a more detailed argument.  Differentiability of $c_W(\lambda)$ implies that (\ref{W_TRANSFORM}) can be computed from calculus: $I_W(z) = \lambda_z z - c_W(\lambda_z)$, where $\lambda_z$ is defined implicitly through $c_W'(\lambda_z)=z$.  As the free energy may not be strictly convex, however, $\lambda_z$ is not necessarily unique.  The set $\{\lambda_z \in \R: c_W'(\lambda_z)=z\}$ is in fact a closed interval $[\lambda_z^{min},\lambda_z^{max}]$, which reduces to a single point only for $z$ corresponding to non-linear portions of $c_W(\lambda)$.  The RHS of (\ref{W_TRANSFORM}) achieves its maximum at every $\lambda_z$ in this interval, including, in particular, at the endpoints.

Let us now consider a sequence $z_n$ converging to $z$.  We then have $\liminf_{n \arrow \infty} \lambda_{z_n} \geq \lambda_z^{min}$ and $\limsup_{n \arrow \infty} \lambda_{z_n} \leq \lambda_z^{max}$, which, by continuity of $c_W(\lambda)$, implies that

\begin{align*}
\limsup_{n \arrow \infty} I_W(z_n) 
& = \limsup_{n \arrow \infty} \lambda_{z_n} z_n - c_W(\lambda_{z_n}) \leq I_W(z) \\
& \leq \liminf_{n \arrow \infty} \lambda_{z_n} z_n - c_W(\lambda_{z_n})
= \liminf_{n \arrow \infty} I_W(z_n),
\end{align*}

\noindent or $\lim_{n \arrow \infty} I_W(z_n) = I_W(z)$.

Finally, recall that $I_W(z)$ is called a {\it good rate function} if all sets of the form $\{z \in \R: I_W(z) \leq a\}$ are compact.  This is implied directly by convexity and continuity.   $\Box$

\bigskip

{\bf Remark 3.6:} Differentiability of the free energy is crucial in the proof of the corollary.  Without it, $I_W(z)$ may be non-convex, with the Legendre-Fenchel transform of $c_W(\lambda)$ yielding only its convex envelope.  In such situations there exists no general method for accessing the rate function.  This breakdown in Legendre duality has been proved \cite{EllisB} to be a necessary and sufficient condition for the nonequivalence of the canonical and microcanonical ensembles in statistical mechanics, whereby mean energies equal to slopes not present in the convex free energy profile (due to a nonanalyticity) cannot be realized in the canonical ensemble.  Even given continuous differetiability of the free energy, however, we cannot improve upon the continuity result for the rate function to conclude differentiability.  The reason is that the free energy may contain linear parts, each of which creates a nonanalyticity in the rate function.  This is why we speak of the two as being related by the Legendre-Fenchel transform instead of the better known Legendre transform, which is defined only between differentiable functions.   

\bigskip

\noindent We now prove our primary result for the action functional.

\bigskip

{\bf Theorem 3.7:} {\it The free energy $c_W^B(\lambda)$ exists, is continuously differentiable $\forall \lambda \in \R$ and satisfies, together with $c_W(\lambda)$, the symmetry relation 

\begin{equation}
c_W(\lambda) = c_W^B(-(1+\lambda)),
\label{C_SYMMETRY}
\end{equation}

\noindent implying the fluctuation theorem 

\begin{equation}
I_W(z) - I_W^B(-z) = -z,
\label{FT}
\end{equation}

\noindent where $I_W^B$ is the continuous and strictly convex good rate function of the action functional under the backward process.}

\bigskip

{\bf Proof:}  Obtaining an explicit representation of the free energy $c_W^B(\lambda)$ is simply a matter of bootstrapping from the work we did in Proposition $3.1$ and Theorem $3.3$.  It is implied by the discussion in Section II that if we reverse time and the orientation of the driving rates $\kij(t)$, then the forward process {\it becomes} the backward process up to boundary conditions.  In particular, the backward action functional $W^B(t_0,t)(\omega)$ equals the heat accumulated by traversing the path $\omega$ {\it backward} from time $t-t_0$ to time $0$.  This immediately implies

\begin{equation*}
\deriv{t_0} u_\lambda^B(t_0,t) = -L_\lambda(t-t_0) \, u_\lambda^B(t_0,t)
\end{equation*}

\noindent for the moment generating function $u_\lambda^B(t_0,t)$, with components $u_\lambda^B(i,t_0,t) = \E_{i,t_0}^B [e^{\lambda W^B(t_0,t)}]$.  Intuitively, increasing $t_0$ still decreases the moment generating function, only in the backward process the increments that are lost correspond to jumps at $t-t_0$.  Were we to recapitulate the proof of Proposition $3.1$, the filtration $\F_t$ used in the conditional expectations would be replaced by $\G_t$, representing equivalently the future of the forward process and the past of the backward one.  Making the change of variables $\tau = t - t_0$, with $\utilde^B(\tau,t) = u_\lambda^B(t_0,t)$, we obtain 

\begin{equation}
\deriv{\tau} \utilde^B(\tau,t) = \L_\lambda(\tau) \, \utilde^B(\tau,t).
\label{BACKWARD3}
\end{equation}

\noindent The solution to this equation can be represented as 
$\utilde^B(\tau,t) = e^{\tau H_\lambda^B} P_\lambda^B(\tau,t) \ones$
for some $T$-periodic $P_\lambda^B$ and real $H_\lambda^B$, whose matrix exponential has strictly positive entries.  Writing the free energy as $c_W^B(\lambda) =$ $\limit{t}{\infty} \frac{1}{t} \, \log \la \mu(\cdot,t), \tilde{u}_\lambda^B(t,t) \ra$, by a chain of equalities similar to (\ref{CHAIN}) and the ensuing argument, $c_W^B(\lambda)$ equals the continuously differentiable Perron-Frobenius eigenvalue of $H_\lambda^B$.  The desired properties for its rate function $I_W^B(z)$ then follow from the arguments in Corollary 3.5.

We now use the symmetry relation $L_{-(1+\lambda)}^* = L_\lambda$ (and $\tilde{L}_{-(1+\lambda)}^* =$ $\Ltilde$) observed from (\ref{L_DEF}) to prove that

\begin{equation}
(H_{-(1+\lambda)}^B)^* = H_\lambda.
\label{H_SYMMETRY}
\end{equation}

\noindent Indeed, as $e^{T H_\lambda^B}$ is a flow operator, mapping solutions of (\ref{BACKWARD3}) $T$ units of time forward (in the direction of increasing $\tau$), irrespective of the current time, it has the representation

\begin{equation*}
e^{T H_\lambda^B} = \T \biggl( \exp \int_\tau^{\tau+T} L_\lambda(s) ds \biggl), 
\quad \forall \, 0 \leq \tau \leq t-T.
\end{equation*}

\noindent Fixing one such $\tau$ and letting $\{s_j^n\}_{0 \leq j \leq n}$ be the sequence of uniform partitions of $[\tau,\tau+T]$ with mesh size $T/n$, we then have

\begin{align*}
e^{T (H_{-(1+\lambda)}^B)^*}
	& = \left( e^{T H_{-(1+\lambda)}^B} \right)^* \\
	& = \limit{n}{\infty} \left\{ \exp \biggl(\frac{T}{n} 
			L_{-(1+\lambda)}(s_{n-1}^n) \biggl) \cdots \exp \biggl(\frac{T}{n} 
			L_{-(1+\lambda)}(s_{1}^n) \biggl) \exp \biggl(\frac{T}{n} 
			L_{-(1+\lambda)}(s_{0}^n) 
			\biggl) \right\}^* \\
	& = \limit{n}{\infty} \exp \biggl(\frac{T}{n} L_\lambda(s_0^n) \biggl)
			\exp \biggl(\frac{T}{n} L_\lambda(s_1^n) \biggl) \cdots 
			\exp \biggl(\frac{T}{n} L_\lambda(s_{n-1}^n) \biggl) \\
	& = \limit{n}{\infty} \exp \biggl(\frac{T}{n} \Ltilde(t-s_0^n) \biggl)
			\exp \biggl(\frac{T}{n} \Ltilde(t-s_1^n) \biggl) \cdots 
			\exp \biggl(\frac{T}{n} \Ltilde(t-s_{n-1}^n) \biggl) \\
	& = \T \biggl( \exp \int_{t-(\tau+T)}^{t-\tau} \Ltilde(s) ds \biggl) \\
	& = e^{T H_\lambda},
\end{align*}

\noindent where the final equality is justified by the fact that $(t-\tau) - [t - (\tau + T)] =$ $T$.  We were also able to interchange the limit and adjoint operators because a sequence of matrices $A_n$ converges component-wise to $A$ iff $A_n^*$ converges to $A^*$. By the uniqueness of matrix logarithms, (\ref{H_SYMMETRY}) holds.

To prove (\ref{C_SYMMETRY}), let $H_\lambda v_\lambda = c_W(\lambda) v_\lambda$, where $c_W(\lambda)$ is the maximal eigenvalue of $H_\lambda$ and $v_\lambda$ is its corresponding positive eigenvector.  By similarity of matrices to their adjoints, $c_W(\lambda)$ is also the maximal eigenvalue of $H_\lambda^*$ and, hence, of $H_{-(1+\lambda)}^B$, via (\ref{H_SYMMETRY}).  But we have already shown $c_W^B({-(1+\lambda)})$ to be maximal for $H_{-(1+\lambda)}^B$, and so by uniqueness of the Perron Frobenius eigenvalue, $c_W(\lambda) = c_W^B(-(1+\lambda))$.

The fluctuation theorem (\ref{FT}) now follows from the G\"{a}rtner-Ellis theorem:

\begin{equation}
\begin{split}
I_W(z) & = \sup_{\lambda \in \R} \{ \lambda z - c_W(\lambda) \} \\
	& = \sup_{\lambda \in \R} \{ \lambda z - c_W^B(-(1+\lambda)) \} \\
	& = \sup_{\lambda \in \R} \{ -(1+\lambda) z - c_W^B(\lambda) \} \\
	& = I_W^B(-z) - z. \quad \Box \\
\end{split}
\label{GARTNER_ELLIS}
\end{equation}

\bigskip

{\bf Remark 3.8:} The careful reader will notice that the $\lambda$ dependence in the relation (\ref{C_SYMMETRY}) that we obtain matches that in the free energy symmetries of Refs. \cite{JiangQianZhang} and \cite{GeJiangQian}, but not Ref. \cite{Lebowitz}.  The reason is one of convention.  The scaled cumulant generating function $e(\lambda) = \limit{t}{\infty} -\frac{1}{t} \log \la e^{-\lambda W(t)} \ra$ in Ref. \cite{Lebowitz} satisfying the symmetry $e(\lambda)=e(1-\lambda)$ is the {\it canonical} free energy function \cite{Touchette}, familiar from statistical mechanics when $t$ is the particle number, $W(t)$ the Hamiltonian of a configuration and $\lambda$ the inverse temperature, but distinct from our free energy function (\ref{FREE_ENERGY}) by the minus signs.  It nevertheless gives rise to the same symmetry for the rate function as ours (in the special case of homogeneous dynamics) via the Legendre-like identity $I_W(z) = \sup_{\lambda \in \R} \{ e(\lambda) - \lambda z \}$ \cite{Lebowitz}.  For an in depth discussion of the relationship between the physics and mathematical notions of entropy and free energy, see Ref. \cite{Touchette}.

\bigskip

{\bf Remark 3.9:} When the generator $A(t)$ is symmetric within each driving period, $I_W^B = I_W$ and our fluctuation theorem (\ref{FT}) reduces to the usual $I_W(z) - I_W(-z) = -z$, which says that the odd part of $I_W$ is linear with slope $-1/2$.  When time-inhomogeneities are involved, however, the fluctuation theorem no longer represents an internal symmetry of a single rate function.  If we interpret it as a formula for computing the large deviations of heat dissipation in the forward process in terms of those in the backward process, then the invariance of (\ref{FT}) with respect to the joint transformation $I_W^B \leftrightarrow I_W$, $z \leftrightarrow -z$, which amounts to reversing the roles of the forward and backward process (recall that $W^B(0,t) = -W(0,t)$), indicates an invariance in the large deviations of heat dissipation with respect to time reversal.

\section{LARGE DEVIATIONS OF ENTROPY PRODUCTION}

We now prove results analogous to the ones above, but with the action functional $W(0,t)$ replaced by the entropy production $S(0,t)$ of the stochastic particle, whose free energies under the forward and backward processes are

\begin{equation}
\label{ENTROPY_FREE_ENERGIES}
c_S(\lambda) = \limit{t}{\infty} \frac{1}{t} \, \log \E_{\pi,0}
	\bigl(e^{\lambda S(0,t)} \bigl) \quad \text{and} \quad
c_S^B(\lambda) = \limit{t}{\infty} \frac{1}{t} \, 
	\log \E_{\mu(\cdot,t),t}^B \bigl( e^{\lambda S^B(0,t)} \bigl).
\end{equation}

\noindent Because the proofs are very similar, we cover only the modifications that must be made.

\bigskip

{\bf Theorem 4.1:} {\it Given the periodicity, continuity, irreducibility and ergodic consistency conditions on $A(t)$, the free energy $c_S(\lambda)$ exists, is continuously differentiable $\forall \lambda \in \R$, and is independent of $\pi$. The time-averaged entropy production $S(0,t)/t$ therefore satisfies a large deviation principle with continuous and strictly convex rate function $I_S(z) = \sup_{\lambda \in \R} \{\lambda z - c_S(\lambda) \}$.} 

\bigskip

{\bf Proof:}  Let $\upsilon_\lambda(t_0,t)$ denote the moment generating function of $S(t_0,t)$, with elements $\upsilon_\lambda(i,t_0,t) = \E_{i,t_0} [e^{\lambda S(t_0,t)}]$.  The derivation of the backward equation we prove for $\upsilon_\lambda$ begins identically to that in Proposition 3.1, except that we replace $w(i,j,t)$ with 

\begin{equation*}
s(i,j,t_0,t) = \log \frac{\kij(t_0)}{\kji(t_0)} + \log \frac{\mu(i,t_0)}{\mu(j,t)}.
\end{equation*}

\noindent These new increments incorporate the boundary terms so that the entropy production becomes an additive process like the action functional, with the representation 

\begin{equation*}
S(t_0,t) = \limit{||\mathcal{P}||}{0} \sum_{i=0}^n s(X_{t_i},X_{t_{i+1}},t_i,t_{i+1}),
\end{equation*}

\noindent where $\mathcal{P} = \{t_i\}_{i=0}^n$ is a partition of $[t_0,t]$.  Starting from the expression (\ref{BACKWARD_LINE}), we have

\begin{align*}
\limit{h}{0+} \upsilon_\lambda(i,t_0-h,t) 
	= & \limit{h}{0+} \sum_{j=1}^N e^{\lambda s(i,j,t_0-h,t_0)} \upsilon_\lambda(j,t_0,t)
		p(i,j,t_0-h,t_0) \\
	= & \limit{h}{0+} \sum_{j \neq i}^N 
		\biggl( \frac{\kij(t_0-h)\mu(i,t_0-h)}{\kji(t_0-h)\mu(j,t_0)} \biggl)^\lambda 
			(\kij(t_0)h + o(h)) \upsilon_\lambda(j,t_0,t) \\
		& + \biggl( \frac{\mu(i,t_0-h)}{\mu(i,t_0)} \biggl)^\lambda
			\bigl(1 - K_i(t_0)h + o(h) \bigl) \upsilon_\lambda(i,t_0,t). \\
\end{align*}

\noindent By the forward evolution equation (\ref{FORWARD}) for $\mu$, 

\begin{align*}
	\biggl( \frac{\mu(i,t_0-h)}{\mu(i,t_0)} \biggl)^\lambda
	& = 1 - h\lambda \frac{\partial_t \mu(i,t_0)}{\mu(i,t_0)} + o(h)  \\
	& = 1 - h\lambda \sum_{j \neq i}^N \kji(t_0) \frac{\mu(j,t_0)}{\mu(i,t_0)}
		+ h\lambda K_i(t_0) + o(h),
\end{align*}

\noindent which we insert into the previous expression to obtain

\begin{align*}
\limit{h}{0+} \upsilon_\lambda(i,t_0-h,t) 
	= & \limit{h}{0+} \sum_{j \neq i}^N 
		\biggl( \frac{\kij(t_0-h)\mu(i,t_0-h)}{\kji(t_0-h)\mu(j,t_0)} \biggl)^\lambda 
			(\kij(t_0)h + o(h)) \upsilon_\lambda(j,t_0,t)  \\
		& + \bigl(1 - K_i(t_0)h + o(h) \bigl) \upsilon_\lambda(i,t_0,t)\\
		& - \upsilon_\lambda(i,t_0,t)  h\lambda
			\sum_{j \neq i}^N \kji(t_0) \frac{\mu(j,t_0)}{\mu(i,t_0)} 
		  + h\lambda K_i(t_0) \upsilon_\lambda(i,t_0,t). \\
\end{align*}
		
\noindent Our component-wise derivative for $\upsilon_\lambda(i,t_0,t)$ then becomes

\begin{align}
\deriv{t_0} \upsilon_\lambda(i,t_0,t) 
= & \limit{h}{0+} \frac{1}{h} [\upsilon_\lambda(i,t_0,t) - \upsilon_\lambda(i,t_0 - h,t)] \notag \\
= & - \biggl[ \sum_{j \neq i}^N \kij(t_0)^{1+\lambda} \kji(t_0)^{-\lambda}
		\biggl( \frac{\mu(i,t_0)}{\mu(j,t_0)} \biggl)^\lambda \upsilon_\lambda(j,t_0,t) 	   \notag \\
\label{UPSILON_DERIV}
	& + \biggl( \lambda \sum_{j \neq i}^N \kji(t_0) \frac{\mu(j,t_0)}{\mu(i,t_0)}
		+ (1 - \lambda)K_i(t_0) \biggl) \upsilon_\lambda(i,t_0,t) \biggl],
\end{align}

\noindent which yields the linear backward equation 
$\deriv{t_0} \upsilon_\lambda(t_0,t) = -M_\lambda(t_0) \upsilon_\lambda(t_0,t)$, where

\begin{equation}
M_\lambda(t_0)_{ij} =
\begin{cases} 
\kij(t_0)^{1+\lambda} \kji(t_0)^{-\lambda}
		\bigl( \frac{\mu(i,t_0)}{\mu(j,t_0)} \bigl)^\lambda, & i \neq j \\
\lambda \sum_{\ell \neq i}^N k_{\ell i}(t_0) \frac{\mu(\ell,t_0)}{\mu(i,t_0)}
		+ (1 - \lambda)K_i(t_0), & i = j \\
\end{cases}
\label{M}
\end{equation}

\noindent The existence of the off-diagonal terms of $M_\lambda(t_0)$ for $t_0 < t^*$ appears problematic here, because we are dividing by $\mu(j,t_0)$, which may be zero.  However, in order for this to be a problem, we must have $\kij(t_0)$ and $\mu(i,t_0)$ both positive.  By continuity of these quantities, they must have been positive on $[t_0 - \delta,t_0]$ for some small $\delta$, ensuring that $\mu(j,t_0) > 0$.  The same argument justifies the finiteness of the diagonal terms.

The rest of the proof is identical to that of Theorem 3.3 and Corollary 3.5, where we make the change of variables $\tau = t - t_0$ and use the Floquet theorem to factor the new solution $\tilde{\upsilon}_\lambda(\tau,t)$ into a periodic matrix and an exponential flow matrix with strictly positive entries.  The free energy $c_S(\lambda)$ is then the principal Floquet eigenvalue of the latter, which by uniqueness is a continuously differentiable function of $\lambda$, implying a large deviation principle with continuous and strictly convex good rate function equal to the Legendre-Fenchel transform of $c_S(\lambda)$.  $\Box$

\bigskip

{\bf Remark 4.2:} Equation (\ref{UPSILON_DERIV}) and definition (\ref{M}) reduce to equation (11) and the subsequent definition of $\ell(\lambda)$ in Ref. \cite{JiangQianZhang} when the time-dependence of our transition rates is dropped.  Conspicuous in the definitions of $M_\lambda(t_0)$ and $\ell(\lambda)$ is that, unlike for $L_\lambda$, the adjoint and $\lambda \arrow -(1 + \lambda)$ transformations are not inverses, implying that the analogue of $H_\lambda$ in the proof above - call it $\hat{H}_\lambda$ - will not satisfy (\ref{H_SYMMETRY}).  This does not invalidate the fluctuation theorem, however, because we merely require that $\hat{H}_\lambda$ and $\hat{H}_{-(1+\lambda)}^*$ have the same maximum eigenvalue, not necessarily be equal themselves.  The fluctuation theorem for entropy production, in fact, is very straightforward.

\bigskip

{\bf Theorem 4.3:} {\it The free energy $c_S^B(\lambda)$ exists, is continuously differentiable $\forall \lambda \in \R$ and satisfies, together with $c_S(\lambda)$, the symmetry relation 

\begin{equation*}
c_S(\lambda) = c_S^B(-(1+\lambda)),
\end{equation*}

\noindent implying the fluctuation theorem 

\begin{equation}
I_S(z) - I_S^B(-z) = -z,
\label{FT2}
\end{equation}

\noindent where $I_S^B$ is the continuous and strictly convex good rate function of the entropy production under the backward process.}

\bigskip

{\bf Proof:}  $\P$ and $\P^B$ are mutually absolutely continuous by the ergodic consistency of $A(t)$ \cite{GeJiang}.  That their Radon-Nikodym derivatives with respect to each other are then reciprocals almost surely is a basic measure-theoretic fact \cite{Folland}. Together, these justify the moment generating function symmetry

\begin{align*}
\E_{\pi,0} \bigl( e^{\lambda S(0,t)} \bigl)
& = \int_\Omega \biggl( \frac{d\P}{d\P^B}(\omega) \biggl)^\lambda
	\P(d\omega) \\
& = \int_\Omega \biggl( \frac{d\P}{d\P^B}(\omega) \biggl)^\lambda
	\frac{d\P}{d\P^B}(\omega) \P^B(d\omega) \\
& = \int_\Omega \biggl( \frac{d\P^B}{d\P}(\omega) \biggl)^{-(1+\lambda)}
	\P^B(d\omega) \\
& = \E_{\mu(\cdot,t),t}^B \bigl( e^{-(1+\lambda) S^B(0,t)} \bigl), 	
\end{align*}

\noindent which is equivalent to the fundamental fluctuation relation of Harris and Sch\"{u}tz \cite{Harris}, cast instead in a quantum Hamiltonian formalism.  We then immediately have the free energy symmetry

\begin{equation*}
c_S(\lambda) 
= \limit{t}{\infty} \frac{1}{t} \log \E_{\pi,0} \bigl( e^{\lambda S(0,t)} \bigl)
= \limit{t}{\infty} \frac{1}{t} \log \E_{\mu(\cdot,t),t}^B 
	\bigl( e^{-(1+\lambda) S^B(0,t)} \bigl)
= c_S^B(-(1+\lambda)),
\end{equation*}

\noindent in which the existence and continuous differentiability of the RHS comes along for free, along with the stated properties of its corresponding rate function.  By the G\"{a}rtner-Ellis theorem and equalities analogous to (\ref{GARTNER_ELLIS}), this implies the fluctuation theorem symmetry.  $\Box$

\bigskip

\noindent Informally, the fluctuation theorem symmetry (\ref{FT2}) yields the following generalization of the second law of thermodynamics:

\begin{equation}
\label{INFORMAL}
\frac{P(S(0,t)/t = z)}{P^B(S^B(0,t)/t = -z)} \sim e^{zt}
\end{equation}

\noindent This has the interpretation that spontaneous positive fluctuations in the time-averaged entropy production of the forward process are exponentially more likely than spontaneous {\it negative} fluctuations in the {\it backward} process, and vice-versa.  Thus, for processes driven by time-dependent protocols, the distributions of entropy production fluctuations in universes in which time moves forward versus backward are interrelated.  Only when the protocol is symmetric within each driving period (i.e., has no temporal orientation) does (\ref{INFORMAL}) reclaim its usual irreversibility interpretation that positive entropy production fluctuations in the forward process are exponentially more likely than negative ones. 

Taking $\Delta z = [z-\epsilon,z+\epsilon]$, $-\Delta z = [-z-\epsilon,-z+\epsilon]$ and $|\Delta z| = \epsilon$, we formalize (\ref{INFORMAL}) as follows.

\bigskip

{\bf Corollary 4.4:} 

\begin{equation*}
\limit{|\Delta z|}{0} \limit{t}{\infty} \frac{1}{t} \log 
\frac{P(S(0,t)/t \in \Delta z)}{P^B(S^B(0,t)/t \in -\Delta z)} = z.
\end{equation*}

{\bf Proof:}  By both the large deviation upper and lower bounds \cite{EllisC}, $\limit{t}{\infty} \frac{1}{t} \log P(S(0,t)/t \in J) = - \inf_{\zeta \in J} I_S(\zeta)$ for any closed, bounded interval $J \subset \R$, and similarly for $S^B(0,t)$.  Therefore, by (\ref{FT2}) and continuity of $I_S$ and $I_S^B$,

\begin{align*}
\limit{|\Delta z|}{0} \limit{t}{\infty} \frac{1}{t} \log 
	\frac{P(S(0,t)/t \in \Delta z)}{P^B(S^B(0,t)/t \in -\Delta z)}
& = \limit{|\Delta z|}{0} - \inf_{\zeta \in \Delta z} I_S(\zeta) 
	+ \inf_{\zeta \in -\Delta z} I_S^B(\zeta) \\
& = -I_S(z) + I_S^B(-z) \\
& = z. \quad \Box
\end{align*} 

\section{TIME-AVERAGED ENTROPY PRODUCTION AND ENTROPY PRODUCTION RATE}

Our final main result is a derivation of the asymptotic time-averaged entropy production and its associated instantaneous entropy production rate.  We begin with the following lemma, which is a well-known large deviations result for countable sequences of random variables \cite{EllisC} and requires only slightly more work in the continous parameter case.

\bigskip

{\bf Lemma 5.1:} {\it The time-averaged entropy production $S(0,t)/t$ converges exponentially and $\P$-a.s. to $c_S'(0)$.}

\bigskip

{\bf Proof:}  By Theorem 4.1, the rate function

\begin{equation}
\label{TRANSFORM}
I_S(z) = \sup_{\lambda \in \R} \{\lambda z - c_S(\lambda) \}
\end{equation}

\noindent is strictly convex.  It is immediate from the definition (\ref{ENTROPY_FREE_ENERGIES}) that $c_S(0)=0$, and so from the inverse transform $c_S(\lambda) = \sup_{z \in \R} \{\lambda z - I_S(z) \}$ we obtain 

\begin{equation*}
0 = c_S(0) = \sup_{z \in \R} \{-I_S(z)\} = - \inf_{z \in \R} \{I_S(z)\},
\end{equation*}

\noindent which, along with strict convexity, implies that $I_S(z)$ attains its minimum and zero uniquely at some $z^*$.  That $z^* = c_S'(0)$ can be seen from the following argument.  Recall from the proof of Corollary 3.5 that $I_S(z) = \lambda_z z - c_S(\lambda_z)$, where $\lambda_z$ is defined implicitly through $c_S'(\lambda_z)=z$.  At the same time, since  $c_S(0)=0$, it is clear that (\ref{TRANSFORM}) is minimized when $\lambda \equiv \lambda_{z^*}=0$, or $z^*=c_S'(0)$.

Exponential convergence of $S(0,t)/t$ to $c_S'(0)$ is now immediate, because for any $\epsilon >0$, the large deviation upper bound

\begin{equation*}
\limsup_{t \arrow \infty} \frac{1}{t} \log 
\P \biggl( \biggl| \frac{S(0,t)}{t} - c_S'(0) \biggl| \geq \epsilon/2 \biggl) \,
\leq - \inf_{|z - c_S'(0)| \geq \epsilon/2} I_S(z) \equiv - \alpha_\epsilon < 0
\end{equation*}

\noindent guarantees that for large $t$, 

\begin{equation}
\label{LDP}
\P \biggl( \biggl| \frac{S(0,t)}{t} - c_S'(0) \biggl| \geq \epsilon/2 \biggl) \, 
	\leq e^{-\frac{\alpha_\epsilon}{2} t}.
\end{equation}

\noindent (The factor $1/2$ multiplying $\epsilon$ will be needed in a triangle inequality argument to follow.)  

Almost sure convergence now follows from several applications of the Borel Cantelli lemma, which says that if the probabilities of a countable sequence of events are summable, then the events happen finitely often almost surely (i.e., only a finite number of them occur).  In this case, we have that the events $|S(0,n)/n - c_S'(0)| \geq \epsilon/2$ occur finitely often, where time has been restricted to the positive integers.  To show that the same holds for general $t$, let $(t_j)_{j \geq 1}$ be any sequence of real numbers tending to $\infty$.  Then 

\begin{align*}
&\P \biggl( \biggl| \frac{S(0,t_j)}{t_j} - \frac{S(0,\tj)}{\tj} \biggl| 
	\, \geq \epsilon/2 \biggl) \\
& \quad \leq \P \biggl( \biggl| \frac{S(0,t_j)}{t_j} - c_S'(0) \biggl| + 
	\biggl|c_S'(0) - \frac{S(0,\tj)}{\tj} \biggl| \, \geq \epsilon/2 \biggl) \\
& \quad \leq \P \biggl( \biggl| \frac{S(0,t_j)}{t_j} - c_S'(0) \biggl| \geq \epsilon/4 \biggl)
	\,+\, \P \biggl( \biggl|\frac{S(0,\tj)}{\tj} - c_S'(0) \biggl| \, \geq \epsilon/4 \biggl).
\end{align*}

\noindent But by an inequality analogous to (\ref{LDP}), these latter probabilities can be bounded by a common exponential for large $t_j$, indicating, by Borel-Cantelli, that $S(0,t_j)/t_j$ and $S(0,\tj)/\tj$ differ by more than $\epsilon/2$ only finitely often (almost surely). The triangle inequality then gives us $|S(0,t_j)/t_j - c_S'(0)| < \epsilon$ for large $t$, precluding, as the $t_j$'s were arbitrary, the possibility of $|S(0,t)/t - c_S'(0)| \geq \epsilon$ for large $t$. This is precisely the definition of almost sure convergence to $c_S'(0)$.   $\Box$

\bigskip

{\bf Theorem 5.2:} {\it The time-averaged entropy production satisfies

\begin{equation}
0 \leq \limit{t}{\infty} \frac{S(0,t)}{t} = \frac{1}{T} \int_0^T e_p(s)ds < \infty,
\label{TIME_AVG_INEQ}
\end{equation}

\noindent where the convergence is exponential and $\P$-a.s., the instantaneous entropy production rate

\begin{equation}
e_p(s) = \frac{1}{2}\sum_{i,j=1}^N (\nu(i,s)\kij(s)-\nu(j,s)\kji(s)) 
			\log \biggl( \frac{\nu(i,s)\kij(s)}{\nu(j,s)\kji(s)} \biggl)
\label{EP_RATE}
\end{equation}

\noindent is defined for $0 \leq s \leq t$, and $\nu(\cdot,t) = \limit{t}{\infty} \mu(\cdot,t)$ is $T$-periodic.}

\bigskip

{\bf Proof:}  By the lemma, we already have convergence in both senses to $c_S'(0)$, so what remains is to evaluate this derivative.  In doing so, we make use of the fact that the free energy is convex and that limits and derivatives commute for sequences of convex functions \cite{EllisC}.

\begin{align*}
{c_S}'(0) & = \deriv{\lambda} \limit{t}{\infty} \frac{1}{t} \log \E_{\pi,0} 
				\bigl( e^{\lambda S(0,t)} \bigl) \Bigl|_{\lambda=0}  \\
	  & = \limit{t}{\infty} \frac{1}{t} \deriv{\lambda} \log \E_{\pi,0} 
				\bigl( e^{\lambda S(0,t)} \bigl) \Bigl|_{\lambda=0}  \\
	  & = \limit{t}{\infty} \frac{1}{t} \E_{\pi,0} S(0,t) \\
	  & = \limit{t}{\infty} \frac{1}{t} \HH(\P,\P^B) \\
\end{align*}

Recall that $\HH(\P,\P^B)$ here is the relative entropy of $\P$ with respect to $\P^B$, which are mutually absolutely continuous by the ergodic consistency criterion on $A$.  By nonnegativity of the relative entropy, we have the left inequality of (\ref{TIME_AVG_INEQ}).  Our goal is now to factor the Radon-Nikodym derivative $d\P/d\P^B$ for finite $t$.  To accomplish this, we note that on the event that the Markov chain jumps $m$ times between times $t_0$ and $t$, its path $\omega$ is characterized by the states $(\sigma_i(\omega))_{i=0}^m$ it visits in sequence and the waiting times $(\tau_{i+1}(\omega) - \tau_i(\omega))_{i=0}^m$ between them.  (Here we have set $\tau_0(\omega) = t_0$ and $\tau_{m+1}(\omega) = t$.)  Because the former are discrete and the latter have a density, as well as the fact that the number of jumps $J(\omega)=J([t_0,t])(\omega)$ over the interval is finite a.s., the path measure $P_{[t_0,t]}$ has density

\begin{equation}
\label{PATH_DENSITY}
f_{[t_0,t]}(\omega) = f_{[t_0,t],J(\omega)}
	(\sigma_0(\omega),\dots,\sigma_{J(\omega)}(\omega),
	 \tau_1(\omega),\dots,\tau_{J(\omega)}(\omega)) 
\end{equation}

\noindent where

\begin{align}
\label{DENSITY}
f_{[t_0,t],m} & (\sigma_0,\dots,\sigma_m,\tau_1,\dots,\tau_m)
  = P(J([t_0,t])=m) \mu(\sigma_0,t_0) \, \times \notag \\
  & \prod_{i=0}^{m-1} \biggl( K_{\sigma_i}(\tau_{i+1})
	\exp \biggl[-\int_{\tau_i}^{\tau_{i+1}} K_{\sigma_i}(s)ds \biggl] 
	\frac{k_{\sigma_i,\sigma_{i+1}}(\tau_{i+1})}
	{K_{\sigma_i}(\tau_{i+1})} \biggl)
	\exp \biggl[-\int_{\tau_m}^t K_{\sigma_m}(s)ds \biggl]
\end{align}

\noindent A similar density $f_{[t_0,t]}^-(\omega)$ exists for the measure $P_{[t_0,t]}^-$.  For notational clarity in what follows, let us write the restriction of a path $\omega$ to an interval $E \subset \R$ as $\omega_E$.  Given an $M \in \Z$ and $t>0$, define the partition variables $s_\ell \equiv s_\ell^{M,t} = \frac{\ell t}{M}$ and scale parameter $h = \frac{t}{M}$.  Recalling that $\P^B(\omega) = \P^-(r(\omega))$, we have by the Markov property

\begin{align*}
& \frac{1}{t} \HH(\P,\P^B)
  = \frac{1}{t} \int_\Omega \log 
	\frac{d\P}{d\P^B}(\omega) \P(d\omega) \\
& \quad = \frac{1}{t} \int_\Omega \log \biggl( 
	\frac{\mu(\omega_0,0)}{\mu^-(r(\omega)_0,0)}
	\prod_{\ell=0}^{M-1} 
	\frac {f_{(s_\ell,s_{\ell+1}]}
			\bigl(\omega_{(s_\ell,s_{\ell+1}]} 
			\bigl| X_{s_\ell} = \omega_{s_\ell} \bigl)}
		  {f_{(s_{\ell},s_{\ell+1}]}^-
			\bigl(r(\omega)_{(s_{\ell},s_{\ell+1}]}
			\bigl| X_{s_{\ell}} = r(\omega)_{s_{\ell}} \bigl)} \biggl) 
	\P(d\omega) \\
& \quad = \frac{1}{t} \sum_{\ell=0}^{M-1} \int_\Omega \log \biggl(
	\frac{\mu(\omega_{s_\ell},s_\ell) 
		  f_{(s_\ell,s_{\ell+1}]}
			\bigl(\omega_{(s_\ell,s_{\ell+1}]} 
			\bigl| X_{s_\ell} = \omega_{s_\ell} \bigl)}			
		 {\mu(\omega_{s_{\ell+1}},s_{\ell+1})
			f_{(s_{M-1-\ell},s_{M-\ell}]}^-
			\bigl(r(\omega)_{(s_{M-1-\ell},s_{M-\ell}]}
			\bigl| X_{s_{M-1-\ell}} = r(\omega)_{s_{M-1-\ell}} \bigl)} \biggl)
	\P(d\omega) \\
\end{align*}

\noindent Here we have rewritten the boundary term 
$\frac{\mu(\omega_0,0)}{\mu^-(r(\omega)_0,0)} = \frac{\mu(\omega_0,0)}{\mu(\omega_t,t)}$
as the telescoping product 
$\prod_{\ell=0}^{M-1} \frac{\mu(\omega_{s_\ell},s_\ell)}{\mu(\omega_{s_{\ell+1}},s_{\ell+1})}$
and reversed the order of the density terms in the denominator.  On the sets

\begin{equation*}
\Omega_{\ell,t,M,i,j} = \{\omega \in \Omega: 
	\omega_{s_\ell}=\omega_{s_\ell -}=i \text{ and } \omega_{s_{\ell+1}}=\omega_{s_{\ell+1} -}=j \},
\end{equation*}

\noindent which partition $\Omega$ up to a set of zero measure (the $s_\ell$ are continuity points of $X_t$ a.s.), the densities on top and botton and the measure integrated against simplify so that we obtain 

\begin{align*}
& \frac{1}{t} \sum_{\ell=0}^{M-1} \sum_{i,j=1}^N \int_{\Omega_{\ell,t,M,i,j}} \log
	\frac{\mu(i,s_\ell) [P(X_{s_{\ell+1}} = j | X_{s_\ell} = i) + o(h)]}
		 {\mu(j,s_{\ell+1}) [P^-(X_{s_{M-\ell}} = i | X_{s_{M-1-\ell}} = j) + o(h)]}
	\P(d\omega) \\
& = \frac{1}{t} \sum_{\ell=0}^{M-1} \sum_{i,j=1}^N \log
	\frac{\mu(i,s_\ell) \kij(s_\ell)h + o(h)}
		 {\mu(j,s_{\ell+1}) \kji^-(s_{M-1-\ell})h + o(h)}
	\P(\Omega_{\ell,t,M,i,j}) \\
& = \frac{1}{M} \sum_{\ell=0}^{M-1} \frac{1}{h} \sum_{i,j=1}^N \log
	\frac{\mu(i,s_\ell) \kij(s_\ell)}
		 {\mu(j,s_{\ell+1}) \kji^-(s_{M-1-\ell})}
	(\mu(i,s_\ell) \kij(s_\ell) h + o(h)) \\
& = \frac{1}{t} \sum_{\ell=0}^{M-1} h \sum_{i,j=1}^N 
	\mu(i,s_\ell) \kij(s_\ell) \log \frac{\mu(i,s_\ell) \kij(s_\ell)} 
										 {\mu(j,s_{\ell+1}) \kji(s_{\ell+1})}. \\
\end{align*}	

\noindent The $o(h)$ correction in the first line is justified by taking $t \downarrow t_0$ in (\ref{DENSITY}) and noting that $P(J([t_0,t_0+h]) = m) = o(h^m)$.  Letting $M \uparrow \infty$ now in the Riemann sum, by continuity of $\mu(\cdot,t)$ and the rates,

\begin{equation}
\frac{1}{t} \HH(\P,\P^B)
= \frac{1}{t} \int_0^t \sum_{i,j=1}^N \mu(i,s)\kij(s) 
		\log \frac{\mu(i,s)\kij(s)}{\mu(j,s)\kji(s)} ds.
\label{PHI}
\end{equation}
		
By Remark 3.4, for all initial distributions $\pi$, $\mu(\cdot,t)$ is asymptotic to a periodic distribution $\nu(\cdot, t)$ equal to the RHS of (\ref{ASYMPTOTIC}), with which we may replace it in the expression above in the limit $t \arrow \infty$.  Let us now take $\phi(s)$ to be the integrand on the RHS of (\ref{PHI}), so that $\phi(s) \sim e_p(s)$.  Given $\epsilon>0$ and choosing an $M$ large enough so that $|\phi(s) - e_p(s)| < \frac{\epsilon}{2}$ for $s>M$, for sufficiently large $t$

\begin{align*}
\frac{1}{t} \int_0^t |\phi(s) - e_p(s)|ds 
& \leq \frac{1}{t} \int_0^M |\phi(s) - e_p(s)|ds + \frac{1}{t} \int_M^t \frac{\epsilon}{2}ds \\
& \leq \frac{\epsilon}{2} + \frac{\epsilon}{2} \biggl( \frac{t-M}{t} \biggl) < \epsilon.
\end{align*}

\noindent We may then finally conclude, by periodicity of $e_p(s)$, that

\begin{align*}
\limit{t}{\infty} \frac{S(t)}{t} 
& = \limit{t}{\infty} \frac{1}{t} \HH(\P,\P^B)
= \limit{t}{\infty} \frac{1}{t} \int_0^t \phi(s)ds \\
& = \limit{t}{\infty} \frac{1}{t} \int_0^t e_p(s)ds
= \frac{1}{T} \int_0^T e_p(s)ds. \quad \Box
\end{align*}

\bigskip

{\bf Remark 5.3:} The result (\ref{TIME_AVG_INEQ}) that we obtain above is the continuous time analogue of Ge et. al.'s complete entropy production rate \cite{GeJiangQian} for discrete time, periodically time-inhomogeneous Markov chains.  The form of $e_p(s)$ itself is a time-dependent generalization of both (33) in Ref. \cite{Gaspard}, where $e_p(s)$ is interpreted as the difference between backward and forward dynamical entropy rates, and (4) in Ref. \cite{JiangQianZhang}.  From a dynamical systems perspective, $\nu(\cdot,t)$ is an attracting limit cycle for $\mu(\cdot,t)$ in the space of distributions on $\{1,\dots,N\}$, representing a sort of periodic steady state.  The mean entropy produced along this cycle is the mean entropy produced by $\mu(\cdot,t)$ in the long time limit.  We see below that when the time dependence of the transition rates is dropped, the limit cycle collapses to a fixed point and $e_p(s)$ reduces to the constant entropy production rate of homogeneous chains.

\bigskip

{\bf Corollary 5.4:} {\it When $\kij(t) \equiv \kij$, $\forall \, 0 \leq i,j \leq N$, 

\begin{equation}
0 \leq \limit{t}{\infty} \frac{S(0,t)}{t} = e_p
\equiv \frac{1}{2}\sum_{i,j=1}^N (\mu(i)\kij-\mu(j)\kji) 
	\log \biggl( \frac{\mu(i)\kij}{\mu(j)\kji} \biggl),
\label{EP}
\end{equation}

\noindent where the convergence is exponential and $\P$-almost sure and $\mu$ is the unique invariant distribution of $X_t$.}

\bigskip

{\bf Proof:}  The existence and uniqueness of $\mu$ are guaranteed by the ergodic theorem.  We then have $\nu(\cdot,t) = \limit{t}{\infty} \mu(\cdot,t) = \mu$, from which the result immediately follows by Theorem 5.2.
\quad $\Box$

\section{BEYOND PERIODIC DRIVING}

The key modeling assumption made in this paper is that the transition rates driving the Markov chain are time-periodic.  Despite the wealth of interesting physical and biological processes characterized by this type of driving, examples of which were given in the introduction, an open question remains as to whether there exists a reasonable set of minimal conditions on the transition rates that are weaker than periodicity but still guarentee existence of an AFT.  By "reasonable", we mean that the conditions be both concise and easily verifiable for a wide range of applications.  It is well known that transient fluctuation theorems make no assumptions about the protocol driving the process, suggesting that the minimal conditions necessary for an AFT to hold truly would be minimal.  An initial guess might be that uniform continuity and boundedness are sufficient, which would prevent the number of jumps of the chain from growing faster than linearly in time as well as preventing pathological behavior due to discontinuous or infinitely rapid driving.  The following result shows that this guess is incorrect. 

\bigskip

{\bf Proposition 6.1:} {\it Given arbitrary constants $0 < \alpha < \beta$, there exists a Markov chain whose transition rates are uniformly continuous and bounded between $\alpha$ and $\beta$, but whose time-averaged entropy production does not satisfy an AFT.}

\bigskip

{\bf Proof:} The plan is to take $0 < \alpha < \beta$ as given and then construct a chain with the properties above.  We begin by choosing distinct constants $\alpha < \kij < \beta$ for $1 \leq i, j \leq N$ such that the generator $A_c = (\kij)$ does not satisfy detailed balance, which of course is always possible.  Note that $A_c$ is irreducible because all its entries are positive, and so it possesses a unique ergodic distribution.  We now choose a number $\gamma > 1$ such that $\alpha < \gamma \kij < \beta$ for all $i$ and $j$, and let $\tau$ be the mixing time of the chain generated by $\gamma A_c$ (the typical time until convergence to its ergodic distribution), which is equal to the reciprocal of the second largest real part of the eigenvalues of $\gamma A_c$.  The generator $A(t)$ of our chain is constructed as follows.  With $t_0 \equiv 0$, we choose $t_1 \gg \tau$ and then iteratively set $t_k = k t_{k-1}$.  $A(t)$ is initially defined to be $A_c$ for $t_{2k} \leq t < t_{2k+1}$ and $\gamma A_c$ for $t_{2k+1} \leq t < t_{2k}$.  We then modify it by smoothing out the discontinuities in any manner that leaves $A(t)$ uniformly continuous in time.  

In order to prove that $S(0,t)/t$ does not satisfy a large deviation property and, hence, an AFT, we show that its free energy $c_S(\lambda)$ fails to converge almost everywhere.  To begin, define $c_S(\lambda, t) = \frac{1}{t} \, \log \E_{\pi,0}(e^{\lambda S(0,t)})$ so that $c_S(\lambda) = \lim_{t \arrow \infty} c_S(\lambda,t)$.  Now let $e_p$ denote the instantaneous entropy production rate of a process generated by $A_c$, which is defined by (\ref{EP}) and must be positive because the rates $\kij$ do not satisfy detailed balance.  Note that $e_p$ is homogeneous of degree $1$ with respect to the rates, so that the entropy production rate of the process generated by $\gamma A_c$ is $\gamma e_p$.  Having defined the $t_k$ to grow such that $t_{k-1}/t_k \arrow 0$, over timescales much longer than the mixing times for $A_c$ and $\gamma A_c$, we see by Corollary 5.4 that 

\begin{equation*}
S(0,t) = 
\begin{cases}
t e_p + o(1), & t_{2k} \leq t < t_{2k+1} \\
t \gamma e_p + o(1), & t_{2k+1} \leq t < t_{2k}, \\
\end{cases}
\end{equation*}

\noindent where $o(1)$ denotes a term that vanishes as $t \arrow \infty$.  This implies that for $\lambda > 0$,

\begin{align*}
\liminf_{t \arrow \infty} c_S(\lambda,t)
& = \lim_{t \arrow \infty} \frac{1}{t} \, 
	\log \E_{\pi,0} \bigl( e^{\lambda (t e_p + o(1))} \bigl) \\
& = \lim_{t \arrow \infty} \frac{1}{t} \, 
	\log \E_{\pi,0} \bigl( e^{\lambda t e_p} \bigl) \\
& < \lim_{t \arrow \infty} \frac{1}{t} \, 
	\log \E_{\pi,0} \bigl( e^{\lambda (t \gamma e_p + o(1))} \bigl) \\
& = \limsup_{t \arrow \infty} c_S(\lambda,t)
\end{align*}

\noindent and hence $c(\lambda)$ does not exist.  The inequality is reversed for $\lambda < 0$, so we see that the free energy only exists for the trivial value $\lambda = 0$.   $\Box$

\bigskip

\noindent Since uniform continuity and boundedness are implied by continuity and periodicity, the preceding result shows that the minimal conditions on the transition rates are closer to those assumed in this paper than one might initially suspect.  What periodicity guarentees but uniform continuity and boundedness do not is that the rates cannot be tuned over arbitrarily long timescales, exactly the loophole we exploited above.  It remains to be seen, however, how this condition can be formulated more precisely and whether there are other conditions must be included in our minimal set.

\begin{acknowledgements}
The authors would like to thank Kenneth Lange and Hua Zhou for helpful discussions.  This work was supported by grants from the NSF (DMS-0349195) and the NIH (K25 AI41935), as well as the VIGRE Graduate Fellowship.
\end{acknowledgements}

{}


\begin{thebibliography}{}

\bibitem{Astumian} R. D. Astumian:
Thermodynamics and Kinetics of a Brownian Motor,
{\it Science}, {\bf 276}:917 (1997).

\bibitem{Bonetto} F. Bonetto, G. Gallavotti, A. Giuliani and F. Zamponi: 
Chaotic Hypothesis, Fluctuation Theorem and Singularities, 
{\it J. Stat. Phys.}, {\bf 123}:39 (2006).

\bibitem{Chen} B. Chen, X. Shen, Y. Li, L. Sun, Z. Yin:
Dynamic theory for the mesoscopic electric circuit,
{\it Phys. Lett.} A, {\bf 335}:103 (2005).

\bibitem{Crooks} G. E. Crooks:
Entropy production fluctuation theorem and the nonequilibrium work relation for free energy differences,
{\it Phys. Rev.} E, {\bf 60}, 2721 (1999).

\bibitem{EllisA} R. S. Ellis:
Large deviations for a general class of random vectors,
{\it Ann. Prob.}, {\bf 12}:1 (1984).

\bibitem{EllisC} R. S. Ellis:
{\it Entropy, Large Deviations, and Statistical Mechanics}. 
Springer-Verlag, New York, (1986).

\bibitem{EllisB} R. S. Ellis, K. Haven, B. Turkington: 
Large deviation principles and complete equivalence and nonequivalence results for pure and mixed ensembles, 
{\it J. Stat. Phys.}, {\bf 101}:999 (2000).

\bibitem{EvansA} D.J. Evans, E.G.D Cohen, and G.P. Morris:
Probability of second law violations in steady flows, 
{\it Phys. Rev. Lett.}, {\bf 71}, 2401 (1993).

\bibitem{EvansB} D. J. Evans and D. J. Searles: 
The Fluctuation Theorem, 
{\it Adv. Phys.}, {\bf 51}:1529 (2002).

\bibitem{Feng} E. H. Feng, G. E. Crooks:
Length of Time's Arrow,
{\it Phys. Rev. Lett.}, {\bf 101}, 090602 (2008).

\bibitem{Folland} G. B. Folland:
{\it Real Analysis}. 
John Wiley \& Sons, New York, (1999).

\bibitem{Gallavotti} G. Gallavotti and E. G. D. Cohen:
Dynamical Ensembles in Stationary States, 
{\it J. Stat. Phys.}, {\bf 80}:931 (1995).

\bibitem{Gammaitoni} L. Gammaitoni, P. H\"anggi, P. Jung and F. Marchesoni:
Stochastic resonance,
{\it Rev. Mod. Phys.} {\bf 70}:223 (1998).

\bibitem{Gaspard} P. Gaspard:
Time-Reversed Dynamical Entropy and Irreversibility in Markovian Random Processes,
{\it J. Stat. Phys.}, {\bf 117}:599 (2004).

\bibitem{GeJiang} H. Ge and D. Jiang: 
The transient fluctuation theorem of sample entropy production for general stochastic processes,
{\it J. Phys.} A, {\bf 40}:F713 (2007).

\bibitem{GeJiangQian} H. Ge, D. Jiang and M. Qian:
A Simple Discrete Model of Brownian Motors: Time-periodic Markov Chains,
{\it J. Stat. Phys.}, {\bf 123}:831 (2006).

\bibitem{Harris} R. J. Harris and G. M. Sch\"{u}tz: 
Fluctuation theorems for stochastic dynamics,
{\it J. Stat. Mech.}, P07020 (2007).

\bibitem{JiangBook} Da-Quan Jiang, Min Qian, Min-Ping Qian:
{\it Mathematical Theory of Nonequilibrium Steady States}.
Springer-Verlag, Berlin, (2004).

\bibitem{JiangQianZhang} D. Jiang, M. Qian and F. Zhang:
Entropy production fluctuations of finite Markov chains,
{\it J. Math. Phys.}, {\bf 4}:4176 (2003).

\bibitem{Joubaud} S. Joubaud, N.B. Garnier, S. Ciliberto: 
Fluctuation theorems for harmonic oscillators,
{\it J. Stat. Mech.}, P09018 (2007).

\bibitem{Kurchan} J. Kurchan: 
Fluctuation theorem for stochastic dynamics,
{\it J. Phys.} A, {\bf 31}:3719 (1998).

\bibitem{KurchanB} J. Kurchan:
A Quantum Fluctuation Theorem,
arXiv:cond-mat/0007360v2 [cond-mat.stat-mech].

\bibitem{Lax} P.D. Lax:
{\it Linear Algebra}.
John Wiley \& Sons, New York, (1997).

\bibitem{Lebowitz} J.L. Lebowitz and H. Spohn: 
A Gallavotti-Cohen Type Symmetry in the Large Deviation Functional for Stochastic Dynamics,
{\it J. Stat. Phys.}, {\bf 95}:333 (1999).

\bibitem{MaesA} C. Maes:
The Fluctuation Theorem as a Gibbs Property,
{\it J. Stat. Phys.}, {\bf 95}:367 (1999).

\bibitem{MaesC} C. Maes and K. Neto\u{c}n\'{y}:
Time-Reversal and Entropy,
{\it J. Stat. Phys.}, {\bf 110}:269 (2003).

\bibitem{MaesB} C. Maes, F. Redig and A. Van Moffaert:
On the definition of entropy production, via examples,
{\it J. Math. Phys.}, {\bf 41}:1528 (2000).

\bibitem{Price} H. Price: 
{\it Time's Arrow and Archimedes' Point}.
Oxford University Press, Oxford, (1996).

\bibitem{Puglisi} A. Puglisi, L. Rondoni and A. Vulpiani: 
Relevance of initial and final conditions for the fluctuation relation in Markov processes,
{\it J. Stat. Mech.}, P08010 (2006).

\bibitem{Rakos} A. R\'akos and R.J. Harris: 
On the range of validity of the fluctuation theorem for stochastic Markovian dynamics,
{\it J. Stat. Mech.}, P05005 (2008).

\bibitem{Renshaw} E. Renshaw: 
{\it Modelling Biological Populations in Space and Time}.
Cambridge University Press, Cambridge, (1991).

\bibitem{Schuler} S. Schuler, T. Speck, C. Tietz, J. Wrachtrup and U. Seifert:
Experimental Test of the Fluctuation Theorem for a Driven Two-Level System with Time-Dependent Rates,
{\it Phys. Rev. Lett.}, {\bf 94}, 180602 (2005).

\bibitem{Seifert} U. Seifert:
Entropy Production along a Stochastic Trajectory and an Integral Fluctuation Theorem,
{\it Phys. Rev. Lett.}, {\bf 95}, 040602 (2005).

\bibitem{Sevick} E. M. Sevick, R. Prabhakar, Stephen R. Williams, Debra J. Searles:
Fluctuation theorems,
{\it Annual Review of Physical Chemistry}, {\bf 59}:603 (2008).

\bibitem{Singh} N. Singh:
Onsager-Machlup Theory and Work Fluctuation Theorem for a Harmonically Driven Brownian Particle,
{\it J. Stat. Phys.}, {\bf 131}:405 (2008).

\bibitem{Sterman} G. Sterman:
{\it An Introduction to Quantum Field Theory}.
Cambridge University Press, Cambridge (1993).

\bibitem{Tietz} C. Tietz, S. Schuler, U. Seifert, and J. Wrachtrup:
Measurement of Stochastic Entropy Production,
{\it Phys. Rev. Lett.}, {\bf 97}, 050602 (2006).

\bibitem{Touchette} H. Touchette:
The large deviation approach to statistical mechanics,
arXiv:0804.0327v1 [cond-mat.stat-mech].

\bibitem{vanZonA} R. van Zon and E.G.D. Cohen:
Extension of the Fluctuation Theorem,
{\it Phys. Rev. Lett.}, {\bf 91}, 110601 (2003).

\bibitem{vanZonB} R. van Zon and E.G.D. Cohen:
Extended heat-fluctuation theorems for a system with deterministic and stochastic forces,
{\it Phys. Rev.} E, {\bf 69}, 056121 (2004).

\bibitem{Verhulst} F. Verhulst: 
{\it Nonlinear differential equations and dynamical systems}.
Springer-Verlag, New York, (1996).

\end{thebibliography}
\end{document}